\documentclass[useAMS,usenatbib]{mn2e}
\usepackage[pdftex]{graphicx}

\title[AGN ionization in companion galaxies]{AGN photoionization of gas in companion galaxies as a probe of AGN radiation in time and direction}
\author[W.C. Keel et al.]
{William C. Keel$^{1,2,3}$\thanks{E-mail: wkeel@ua.edu},   Vardha N. Bennert$^{4}$, Anna Pancoast$^{5,6,7}$,
\newauthor  Chelsea E. Harris$^{5,8}$, Anna Nierenberg$^{5,9,10}$,  S. Drew Chojnowski$^{11}$,  
\newauthor Alexei V. Moiseev$^{12,13,14}$, Dmitry V. Oparin$^{12}$,
Chris J. Lintott$^{15,16}$, \newauthor  Kevin Schawinski$^{17}$, Graham Mitchell$^{18}$, 
and Claude Cornen$^{18}$\\
$^{1}$Department of Physics and Astronomy, University of Alabama, Box 870324, Tuscaloosa, AL 35487, USA\\
$^{2}$Visiting Astronomer, Kitt Peak National Observatory, operated by AURA, Inc., under contract to the US National\\ Science Foundation.\\
$^{3}$SARA Observatory\\
$^{4}$Department of Physics, California Polytechnic State University,  San Luis Obispo, CA 93407, USA\\
$^{5}$Department of Physics, University of California, Santa Barbara, CA 93106 USA\\
$^{6}$Harvard-Smithsonian Center for Astrophysics, 60 Garden St., Cambridge, MA 02138 USA\\
$^{7}$Einstein Fellow\\
$^{8}$Department of Energy Computational Science Graduate Fellow, Department of Astronomy, University of California, Berkeley, CA 94720\\
$^{9}$Center for Cosmology and AstroParticle Physics, 191 West Woodruff Avenue, The Ohio State University, Columbus, OH 43204, USA\\
$^{10}$CCAPP Fellow\\
$^{11}$Department of Astronomy, New Mexico State University, Las Cruces, NM  11001, USA\\
$^{12}$Special Astrophysical Observatory, Russian Academy of Sciences, Nizhny Arkhyz 369167, Russia\\
$^{13}$Lomonosov Moscow State University, Sternberg Astronomical Institute, Universitetsky pr. 13, Moscow 119234, Russia\\
$^{14}$Space Research Institute, Russian Academy of Sciences, Profsoyuznaya ul. 84/32, Moscow 117997, Russia\\
$^{15}$Astrophysics, Oxford University 		
$^{16}$Adler Planetarium, 1300 S. Lakeshore Drive, Chicago, IL 60605\\
$^{17}$Department of Physics, ETH Z\"urich, Switzerland			
$^{18}$Galaxy Zoo\\
}

\begin{document}

\date{}

\pagerange{\pageref{firstpage}--\pageref{lastpage}} \pubyear{2018}

\maketitle

\label{firstpage}

\begin{abstract}
We consider AGN photoionization of gas in companion galaxies (cross-ionization)
as a way to sample the intensity of AGN radiation in both direction and time, independent of the gas properties of the
AGN host galaxies. From an initial set of 212 AGN+companion systems, identified with the help of Galaxy Zoo participants,
we obtained long-slit optical spectra of 32 pairs which were {\it a priori} likely to show cross-ionization based on
projected separation or angular extent of the companion. From emission-line ratios, 10 of these systems are candidates for 
cross-ionization, roughly the fraction expected if most AGN have ionization cones with 70$^\circ$ opening angles.
Among these, Was 49 remains the strongest nearby candidate.
NGC 5278/9 and UGC 6081 are dual-AGN systems with tidal debris, complicating identification of cross-ionization.
The two weak AGN in the NGC 5278/9  system ionize gas filaments to a projected radius 14 kpc  from each galaxy. In UGC 6081, 
an irregular high-ionization emission region encompasses both AGN, extending more than 15 kpc 
from each. The observed AGN companion galaxies with and without signs of external AGN photoionization have similar distributions in 
estimated incident AGN flux, suggesting that geometry of escaping radiation or long-term variability control this facet of the AGN environment.  
This parallels conclusions for luminous QSOs based on the proximity effect 
among Lyman $\alpha$ absorbers. In some galaxies, mismatch between spectroscopic classifications in the common BPT diagram and
the intensity of weaker He II and [Ne V] emission lines highlights the limits of common classifications in low-metallicity environments.

\end{abstract}

\begin{keywords}
galaxies: Seyfert --- galaxies: ISM --- galaxies: active
\end{keywords}

\section{Introduction}

Important aspects of active galactic nuclei (AGN) can be studied through  the properties of surrounding gas on scales up to kiloparsecs,
providing information otherwise inaccessible due to the small size of the central regions or the timescales involved. This is especially
true for AGN with extended emission-line regions (EELRs), which span tens of
kiloparsecs in radial distance and often occur over broad angular ranges about the
central source (\citealt{Stockton}). The frequent occurrence of (double) ionized clouds with triangular shapes (``ionization cones") 
was a key piece of evidence for 
some form of geometric unification involving Seyfert galaxies of types 1 and 2 \citep{ski93}. Recently, the discovery of the 45-kpc cloud
known as Hanny's Voorwerp (\citealt{Lintott2009}, \citealt{Kevin2010}, \citealt{HVHST}, \citealt{Kevin2015}, \citealt{Sartori}) 
highlighted what had earlier been an abstract possibility: using EELRs to trace the history of radiation 
from AGN across the light-travel times to distant gas clouds. In this
object, the gas is so highly ionized that it must be illuminated by a luminous quasar, 2 orders of magnitude brighter than 
the observed galaxy nucleus; a former quasar has faded dramatically within the
$\approx 10^5$-year light-travel time between the central AGN and the cloud edge. A tailored search by Galaxy Zoo participants 
led to confirmation  of 19 less luminous
examples (\citealt{mnras2012}), while later studies have identified analogous candidates
for fading AGN detected via such ionization echoes at both lower and
higher luminosity (\citealt{greenbeans}, \citealt{ngc7252}, \citealt{m51cloud}).

The use of EELR gas to determine properties of the AGN is restricted
by the distribution of the gas around 
the AGN host galaxy. The escape of radiation may suffer from biases due both to the relative orientations of the disc 
and AGN absorbing torus (whose axes are essentially uncorrelated; \citealt{Schmitt})
and to obscuration from dust in the galaxy disc; there must be distant gas in directions where the AGN radiation is not blocked by material in the host. Indeed, in Hanny's Voorwerp
and its fainter Galaxy Zoo analogs, the EELRs are part of tidal debris 
\citep{Keel2015}. In this study, we explore a kind of EELR which is not directly related to the AGN host - gas in companion galaxies.
EELRs of this kind
would not suffer the same biases in illumination, and would
also not have the kind of radial density structure which might be found in
gas associated with the AGN host itself; the distribution of gas in the companion would still control where it could be ionized, but now in a way
not correlated with the AGN's structure. These kinds of decoupling from the AGN host structure might allow more secure
measurements of the angular pattern of escaping radiation, and its
time history, than are possible from ionization gas within the AGN host itself.
While the details of triggering AGN by galaxy interactions remains contentious,
many AGN are certainly located in close (and often interacting) galaxy pairs, so we can
use companion galaxies as screens to view the emerging 
radiation from their active neighbors, which for convenience we will call cross-ionization.

The conditions needed for cross-ionization in galaxy pairs to occur
were set out as early as \cite{AVF82}, and have been briefly
considered again in connection with growing interest in dual-AGN systems \citep{Liu}.
In spectra with large apertures, cross-ionization could mimic
a relatively low-luminosity AGN (as was noted by \citealt{Liu}), while
ionized gas from a companion could generate velocity asymmetries in survey spectra
of the kind studied by \cite{Comerford14}. One of the
best candidates, Was 49ab, was observed in detail by \cite{Moran}
who discussed the possibility of cross-ionization.

Parallel scientific issues - lifetime of AGN luminous episodes and the solid angle over which ionizing radiation escapes -
have been addressed for QSOs by seeking changes in the ionization level implied by
absorption lines from dense parts of the intergalactic medium (IGM), above all the
H I and He II Lyman $\alpha$ forests \citep{FurlanettoLidz}. A local source of hard radiation, above the mean
background at a QSO's redshift, decreases the strength of these absorption features
within a radius depending on the local AGN flux. This has been measured as the
proximity effect - originally in the spectrum of the QSOs themselves, with redshift mapping to distance in front of the AGN, and 
later in the transverse proximity effect 
(\citealt{Adelberger2004}, \citealt{KirkmanTytler}, \citealt{Goncalves}, \citealt{Schirber}), as the line of sight to
the AGN observed passes close to a foreground AGN. Mismatch between these two
kinds of proximity effect could result from episodic luminous AGN phases 
\citep{VisbalCroft} or anisotropic escape of the ionizing radiation.

Our approach complements studies of both proximity effects; close companion galaxies
can probe smaller separations and lower AGN luminosities than are feasible with the usually
statistical nature of absorption-line studies. We rely
on an analogous process: detection of ionization in excess of what sources 
other than the AGN can produce, in this case through changes in emission-line ratios as the
AGN contribution becomes comparable to, or exceeds, that from star formation in a
companion galaxy. 

We have conducted an extensive search for AGN cross-ionization candidates. In
section 2 we describe our sample selection, with participation from Galaxy 
Zoo volunteers, the choice of observational priorities among these based on physical properties, 
narrowband imaging of a subset, and 
spectroscopic observations. In section 3  we identify a set of candidates for
cross-ionization between AGN and companion galaxies, as well as noting instances both of AGN potentially missed in
the most common diagnostics due to low gas metallicity, and of close companions showing no influence by luminous AGN. 
Finally, we compare our subsets of AGN companions with and without spectroscopic evidence of cross-ionization. In quoting luminosities and
sizes, we adopt $\rm H_0= 70$ km s$^{-1}$ Mpc$^{-1}$ and flat geometry.

\section{Sample selection and observations}

\subsection{Sample criteria}

For detectable cross-ionization of the ISM in a companion galaxy,
both the impinging intensity of ionizing radiation from the AGN and the chances 
of the companion intercepting the pattern of emerging radiation
(i.e. ionization cones) are relevant. Our
criteria are shown schematically in Fig. \ref{fig-schematic}. Here,
$r$ is the projected distance from the AGN to the center of the
companion, and $\theta$ is the angle the companion subtends in 
projection around the AGN. These are largely independent criteria (although for constant
galaxy size, larger $\theta$ occurs at small $r$),
and we consider both in selecting the most promising targets. Galaxy size is not well-defined,
particularly when images overlap so that standard photometric procedures
may not separate galaxies properly; 
operationally we take the extent detected in composite images from the
Sloan Digital Sky Survey (SDSS; \citealt{SDSSDR7}, \citealt{SDSSDR12})
for a consistent way to compare different systems. Examination of a variety of our systems shows that
this extends to a projected distance from the nucleus roughly equal to the SDSS quantity {\it petroR90\_r}, and
about 1.3 times the Petrosian radius {\it petroRad\_r}. 

Observed ionization cones have projected full opening angles of typically $70\degr$ (\citealt{Wilson}, \citealt{mnras2012}), 
half-angle $\approx 35\degr$. Ignoring projection effects which are poorly-known for most objects, this implies
that ionizing radiation emerges over a fraction ($1 - \cos 35\degr = 0.18$) of the sky around a typical AGN, and suggests that we
could expect roughly this fraction of companions to show cross-ionization if they are close enough for the
incident AGN radiation to be suitably intense. This is a very crude estimate, since there is such a wide range of
observed opening angles and projection factors will enter, but helps to guide our expectations.

Both $r$ and $\theta$ are subject to projection effects due to the (essentially unknown)
line-of-sight separation between the galaxy centers. Since there is no control over these factors in sample
selection, we simply take the observed quantities. However, it may be of interest to estimate the expected projection factors
for galaxy separation and (hence in ambient AGN flux predicted based on this separation) on companion galaxies. 
For this purpose, we carried out a simple Monte Carlo 
simulation based the slope $\zeta$ of the galaxy-galaxy correlation function slope
(which is consistent across the separations of  galaxy pairs and groups), using a cut in projected radius (whose value is arbitrary
since the correlation function is a power law) to model the selection of a sample on projected separation. Using 
$10^4$ Monte Carlo trials for pairs of points separating according to $\zeta \propto r^{-1.6}$, the mean projection effect is 
poorly defined because of the long tail of widely-separated galaxies, but the median projection factor between projected and physical 
radii is stable at 1.52 (decreasing the inferred intensity of incident AGN radiation by a factor 2.3). 

Ideally, we would seek companion galaxies with the rare combination of being rich in neutral gas to trace ionizing radiation, and poor
in ongoing star formation which dilutes the emission lines of AGN-ionized gas. A selection on companion optical colours could eliminate
``red and dead" systems on the red sequence which are very gas-poor, but this could also in principle eliminate some galaxies with
enough gas for observable line emission if they have low rates of star formation (or, for example, H I clouds on the periphery of quiescent galaxies).

When both galaxy redshifts are known, we can filter many line-of-sight projections from candidates for genuine companion galaxies. We required
$ \mid \Delta cz \mid < 400$ km $^{-1}$ to reject line-of-sight companion that are not physically associated; 
as it turned out, a smaller $\Delta v$ threshold
would have given virtually the same selection. Among all pairs we observed spectroscopically,
the median $\mid \Delta cz \mid$ is 93 kms$^{-1}$. All objects we observed with tidal disturbances have component redshifts matching within
our selection threshold.


\subsection{Sample construction}

We compiled a finding list of AGN with companion galaxies based on the redshift and geometric criteria in section 2.1,
largely through the efforts of
volunteer participants in the Galaxy Zoo project \citep{Lintott2008}.
A post on the project forum setting out the desired kinds of
galaxy pairs led to responses beginning both from objects seen in the
normal course of classification for Galaxy Zoo, and from SQL queries of 
the SDSS photometric and spectroscopic catalogs\footnote{http://www.galaxyzooforum.org/index.php?topic=279841.210}. 
This initial query selected object pairs projected within 15\arcsec  
with both redshifts in SDSS DR8, where one
had an AGN spectroscopic class and the other had a non-AGN galaxy spectroscopic class. The cutoff in projected
separation corresponds to 17.5 kpc at the median sample redshift $z=0.060$.
To this we added additional pairs found by Galaxy Zoo volunteers which
satisfied the same criteria except for not having both redshifts available from the SDSS, or fulfilling the linear separation criterion at lower redshifts. 
In some cases the missing redshift
had been measured and was available from sources referenced in NED, and in others, tidal structure made physical association
between the galaxies virtually certain (which was confirmed by our spectroscopy). Similarly, we also included nearby systems passing these tests, previously known outside the SDSS imaging region (i.e. \citealt{Keel1996} building on the catalog by \citealt{Lipovetskii}):
NGC 2992, NGC 6786, Kaz 63, Kaz 199.

This set of 
pairs was further refined by our inspection of
spectra to confirm the clear presence of a spectroscopic AGN.
We required a Seyfert nucleus; originally we started compiling LINER AGN as well,
but as their inferred ionizing luminosities all fell below our threshold for observation,  we ceased collecting
them for this program. When only the AGN redshift is
known, tidal distortion was taken as secondary evidence that two
galaxies are physically associated. These factors led us to a
finding list of 212 AGN/companion pairs (Appendix A), incorporating Galaxy Zoo forum postings
between 16 January 2012 and 15 March 2014.

\begin{figure} 
\includegraphics[width=65.mm,angle=0]{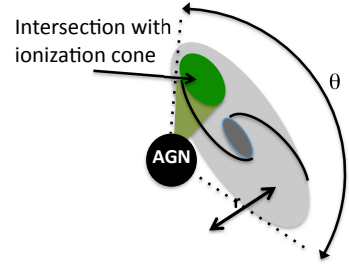} 
\caption{Schematic diagram of an AGN illuminating a large companion galaxy, showing the
selection parameters $r$ and $\theta$. The projected separation $r$, combined with the AGN
luminosity, predicts the strength of ionization expected at the companion, while its
angular extent $\theta$ with respect to the AGN predicts the probability of ionization cones
actually intersecting the companion galaxy. Both $r$ and $\theta$ are subject to projection effects.}
 \label{fig-schematic}
\end{figure}

From this finding list of 212 AGN/companion pairs, we used large $\theta$ and small $r$ as in Fig. \ref{fig-schematic}
to prioritize them for spectroscopic observations. We estimate the ionizing luminosity based on the nuclear [O III]
flux. We use [O III] rather than H$\alpha$ or H$\beta$, since the broad-line components can suffer obscuration in at least some 
Sy 2 nuclei; the [O III] luminosity is a nearly isotropic luminosity indicator based on its relation to mid- and far-IR selection
(e.g. \citealt{Keel1994}). The [O III] fluxes are taken mainly from the SDSS spectra, augmented where necessary by published data from
other sources. We used these values only in a relative sense here, for ranking priorities; a conversion to absolute values is needed to predict 
individual values and ask what threshold in ionizing flux should lead to observable ionization consequences. For this, we used the conversion 
between [O III] and bolometric luminosity from \cite{Heckman2004}, namely 
a ratio of 3500 between bolometric and observed [O III] output, and the mean radio-quiet
spectral energy distribution (SED) from the QED project \citep{QED}, where the ionizing luminosity is about 0.14 of the bolometric value.
These give predictions of the incident ionizing flux F$_{ion}$ on companion galaxies
from the AGN [O III] flux and companion geometry, which can suggest
likely ionization parameters for a particular ISM density. This estimate applies to the projected
location of the companion nucleus; in some systems the range between near and far sides of a companion disc is very large.
The diffuse phase of the ISM is important here; at its typically low
density, relatively few AGN photons can shift its ionization balance, in contrast
to H II regions with higher densities of both particles and stellar ionizing photons. The
contributions of these phases differ considerably among galaxies.  For this reason, we 
use the expected ionizing flux from AGN at companion locations only as a relative guide to 
prioritize our observations.

For each spectroscopic observing session, we selected accessible pairs with either the highest
predicted ionizing flux at the companion, or close pairs with the largest projected
angle $\theta$ spanned by the companion (translated to expected solid angle
spanned by the companion as seen from the AGN). 
Selection for highest incident ionizing flux
will restrict this study to luminous AGN in galaxy pairs; these will be a small subset, given
evidence that AGN in dense environments (including close pairs) are predominantly of low luminosity (\citealt{Martinez2010}, \citealt{Bitsakis2015}),
with additional differences between interacting pairs and dense compact groups \citep{Sabater2013}. Furthermore,
AGN which remain at low luminosity for long periods are unlikely to be able to produce radiatively-driven outflows which
could contribute to extended emission \citep{ElitzurShlosman}.

When the companion spectrum is available (from
SDSS or other sources as listed by NED), we omitted those with strong line emission from star formation,
since weaker contributions from AGN-ionized gas become progressively more
difficult to distinguish in composite systems. Previous dual-AGN candidates were retained in our list,
since some of these may be instances of cross-ionization where the
nucleus and surrounding ISM are blended in the SDSS spectrum \citep{Liu}.
We note in discussing individual candidates below where a second AGN is clearly present,
and how its luminosity affects the likelihood of cross-ionization from the brighter AGN.

Mkn 177 and its close, compact companion object fit our selection criteria, but 
 \cite{Koss2014} find no resolved line emission around the putative AGN,
whose long-term variability suggests that it may instead be a very luminous, unusual variable star, or more speculatively,
an ejected supermassive black hole leaving the site of interaction.

\subsection{Observations: long-slit spectra}

Cross-ionization should give spectroscopic signatures such as extended ionized gas with AGN-like line ratios across the companion, and
ionization behavior not symmetric about the companion nucleus. We have carried out a spectroscopic study of candidate AGN
in galaxy pairs to isolate a subset with these features.

We obtained spectra of 32 candidate pairs, some multiple times for confirmation or at different position angles, using the Kast
double spectrograph \citep{Kast} at the 3-m Shane telescope of Lick Observatory during
13 nights from 2013--2015. For each session, the D46 dichroic beamsplitter
separated light into blue and red optical paths, with a nominal split centered at 4600 \AA. The wavelength
settings were roughly 3400-4600 \AA\ in the blue, and 4600-7400 \AA\ on the red side. The slit width was 2.0 arcseconds. Flux calibration used observations of 1-3 standard stars per night.; clouds prohibited observing standards on 23 April 2015, so we used 
the response curve derived for 26 April, and scaled line fluxes in the case of SDSS 1354+1327
to match earlier data where the slits crossed at the nucleus. Observations on each night are listed in Table \ref{tbl-setups}, and descriptions of the spectral characteristics of spatially resolved, extended line emission are
given in Table \ref{tbl-spectra}. The initial slit orientation crossed both galaxy nuclei, unless SDSS images suggested
strong off-nuclear line emission (in tidal tails) via unusual $gri$ colours, with further slit orientations in some cases when
suggested by system structure or results of a first observation. The data reduction pathway includes bias subtraction, flat-field correction,
resampling to linear wavelength scales, sky subtraction, and flux calibration using nightly standard-star data; these were
done using tasks from the IRAF \citep{Tody1986} {\it twodspec} package.

The NGC 5278/9 system was also  observed  on 10 February 2018 in the long-slit  mode of  the SCORPIO-2 multi-mode focal reducer 
\citep{AM2011}  at the prime focus of the 6-m telescope (BTA) of the Special Astrophysical Observatory, Russian Academy of Sciences. The  
slit  width was  $1$\arcsec, and the spectral coverage was about 3500--7200 \AA\, with a typical resolution $\sim5$ \AA\ . 
Slit positions and orientations were
selected to include some of the near-radial emission-line features outside the galaxy discs, through NGC 5278 at PA 100$^\circ$
and NGC 5279 at PA 91$^\circ$. Each slit position was observed for 3600 seconds. After standard reduction, 
the BTA spectra were flux-calibrated using a standard star observed on the same night. These data were obtained using $1 \times 2$-pixel binning during readout, for a scale of 0.356\arcsec per pixel along the slit, which we summed by a further factor of 3 to get sets of line measures every 1\farcs 07 along the slit.
This provides enough surface-brightness sensitivity to detect He II $\lambda 4686$ emission in some
of the extended features at levels indicating AGN photoionization, independent of issues with the stronger line ratios at low metallicity.

We estimate errors on the emission-line ratios measured in our spectra using the same Gaussian-fitting procedure
as used to derive their radial velocity and fluxes, the deblending option in the IRAF routine {\it splot}. This was done
by adding an emission-line position at an arbitrary wavelength where no line is expected, in the red and blue regions, and
adopting the flux statistics of the resulting fictitious lines as the error distribution. We examined the behavior of this error
distribution with local flux, finding no significant trend (that is, the effects of Poisson noise from the galaxy continuum are weak)
in the Lick spectra. The deeper BTA spectra do show an error increasing with line flux, so we include this factor for the BTA spectra. 
These estimated errors are propagated into the emission-line ratios used to infer ionizing mechanisms.

To assess the dominant ionizing mechanism in our spectra, we rely on the approach
pioneered by \cite{BPT}, widely known as BPT diagrams. In particular, star-forming objects and
AGN are generally well separated in a diagram comparing the ratios of strong emission lines
[O III] $\lambda 5007$/H$\beta$ and [N II] $\lambda 6583$/H$\alpha$. Specifically,
we use the updated boundaries specified by \cite{Kewley2006}. This single diagram,
while relying on the spectral lines most easily detected in the optical, can become ambiguous
at low metal abundances (as in some of the EELRs examined by \citealt{mnras2012}). 
He II or [Ne V], when seen over large areas, are virtually certain indicators of photoionization by
AGN radiation, but our data are only rarely sensitive enough to reveal these in the outskirts of the galaxies. 
Further, the wavelength range around He II $\lambda 4686$ at redshifts $z=0.02 - 0.04$ (typical of the bright objects in this 
sample) is compromised by a night-sky emission line in spectra from Lick Observatory.
 
Of the 32 systems we observed, 10 show evidence of extended AGN-ionized gas in companion
galaxies or tidal tails.
The long-slit results for the 10 objects with possible AGN ionization in the companion galaxies (or tidal tails)
are presented in Fig. \ref{fig-longslit}. Spectral properties were evaluated using Gaussian fits
to 2-pixel (1.5\arcsec, for Lick data) or 3-pixel (1\farcs 07, for BTA data) sums along the slit. The top panels  show
the slit locations in these systems superimposed on composite SDSS $gri$ images. The
middle panel for each spectrum shows the 
redshift behavior (heliocentric $cz$) along the slit as well as intensity slices in [O III], H$\alpha$, and their adjacent continuum regions.  
The bottom panels show
the strong-line BPT diagram with the star-forming/AGN boundary from \citealt{Kewley2006};
the colours of points match between the velocity plots and BPT diagrams to show where the
AGN-ionized regions occur along the slit. 
The effects of underlying stellar absorption drive us to estimate H$\beta$ from H$\alpha$,
since the absorption correction is both smaller and less variable with stellar population for H$\alpha$. The greater strength of
H$\alpha$ further reduces this effect, and the equivalent widths are
small enough in many instances to make the uncertain absorption correction relatively large in spectra with
insufficient signal-to-noise to model the local stellar populations.
We therefore estimate the flux of  H$\beta$ 
from that of (narrow-line components of) H$\alpha$ and a Balmer decrement of 2.85.
A key virtue of the BPT diagram is that it is almost completely reddening-independent. However,
using H$\alpha$ data to estimate H$\beta$, as we do, introduces a reddening dependence,
in the sense that [O III]/H$\beta$ will be underestimated in reddened systems, since the standard
Balmer decrement will lead to an overestimate of the H$\beta$ flux. We can tolerate this effect,
since in all parts of the strong-line BPT diagram, reddening correction will move
data  points further into the AGN region. This approach is thus conservative with regard to  
identifying AGN-photonized regions.

The strong-line BPT diagrams, while allowing examination of the most extensive areas,
may be misleading when low-metallicity gas is involved, such as is encountered in tidal debris and the outer discs of spirals
(\citealt{Groves2006}, \citealt{Kawasaki2017}). 
AGN photoionization
models with varying abundances (oxygen and nitrogen having the strongest signatures) show that nitrogen abundances
below solar can move AGN-ionized gas into parts of the usual BPT diagrams occupied by 
star-forming regions (\citealt{StorchiBergmann}, \citealt{Castro}); in particular, [N II] and [S II]
weaken at fixed [O III]/H$\beta$. So, while AGN-ionized gas may be securely identified
if its strong-line ratios place it in the ``AGN" region of the BPT diagrams, lower-metallicity
counterparts may require additional information for secure classification.
Significant He II or [Ne V] emission can play such a role, but these lines are comparatively
weak in AGN narrow-line regions, and would not be detected outside the nuclei in most of our Lick
spectra.
The [O I] $\lambda 6300$ line is 
detected in some of our spectra and can help resolve this ambiguity, since  this line is very weak
in normal H II regions.

An additional ionizing process - shocks - is found to be important around the jets and in high-velocity
outflows of AGN, particularly radio-loud AGN. Recognizing their role has proven to be subtle from only
the strong optical emission lines. When the \cite{BPT} classification was initially formulated, what became
known as LINERs were broadly identified as shock-ionized. As additional data have accumulated, showing that
some LINERs are indeed photoionized by AGN (at lower ionization parameters than Seyfert nuclei) while others
represent shocked gas in merging systems, and still other are photoionized by hot evolved stars (\citealt{Stasinska}, \citealt{Johansson}),  
integrated strong-line ratios by themselves are seen to give little discrimination on this question. Weaker
lines, especially [O III] $\lambda 4363$ and [O I] $\lambda 6300$, provide more diagnostic power (but are not uniformly detected in
the extended clouds in our data). On the broader question of separating LINERs from Seyfert nuclei, 
\cite{Kewley2006} find that  either [S II]/H$\alpha$ or [O I]/H$\alpha$ gives more consistent results than
[N II]/H$\alpha$ when compared to [O III]/H$\beta$ (the ``standard" strong-line diagram). However, especially
at lower metallicity, these more diagnostic lines become weak and their use becomes less powerful. Pending
data of higher signal-to-noise ratio, we note that important shock contributions around AGN are largely associated
with radio-loud objects, which are relatively rare both in the overall AGN population and our sample.

In addition to line ratios, one might often (although not always) expect kinematic disturbances or broad lines
where shock ionization is important. In the extended emission regions in our sample, we do not see any lines
with the typical FWHM $ > 300$ km s$^{-1}$ or velocity jumps of similar amplitude that are
comparable to the shock velocities needed to give the observed excitation level in [O III]/H$\beta$.

\begin{table*}
 \centering
  \caption{Spectroscopic observations}
  \begin{tabular}{@{}ll@{}}
  \hline
 UT Date &  Galaxies observed \\
 \hline
Jan 15 2013	& 0057+0120,0848+3515\\
Jan 16 2013	&  0029+0010,0848+3515, 1214+2931\\ 
Jan 17 2013	&  0029+0010 compn,0757+3511,0848+3515,1200+147\\
Jan 18 2013	& 0057+0120,0841+0101,0905+3237,1010+0612,1352+2528\\
Mar  10 2013	&  0838+0407,0847+3445,0904+5536,1138+1412,1201-0153,1303-0306,1354+1327\\
Mar 11 2013	&  0904+5536,1213+5138,1243+3738,1342+1839,1347+1734,1354+1327\\
Mar 12 2013	&  1050+2329,1219+1326, 1414-0000\\
Mar 13 2013	&  1042+0502,1101+1017,1132+5257,1142+3251, Kaz 199\\
Feb 4  2014	&  0839+4707, UGC 6081, NGC 5279\\
Feb 5  2014	&  0838+0407, UGC 6081,NGC 5279\\
23 April 2015    &  0848+3515, 1354+1327, NGC 3341, Was 49, NGC 5279\\
24 April 2015 & NGC 3341, 1354+1327 \\
26 April 2015 &  1201-0153, NGC 3341, NGC 5279, UGC 6081, Was 49 \\
\hline
\end{tabular}
\label{tbl-setups}
\end{table*}

\begin{table*}
 \centering
  \caption{Spectroscopic summary}
  \begin{tabular}{@{}lcccl@{}}
  \hline
AGN    & Type & $z$ & PA$^\circ$ & Description \\
\hline
\multispan5 Candidate cross-ionization systems \hfil \\
 & & & & \\
SDSS J002944.89+001011.1& Sy 2 & 0.0598 &   49 & Line emission extended across pair. Compn [O III]/H$\beta$ drops to$ ~\approx 2.5$ (UM 246)\\
SDSS J005754.03+012013.8 &  Sy 2 & 0.0567 & 168 & UM 293. Extended high-ionization emission from AGN to companion. \\
SDSS J083902.96+470756.3 & Sy 2  & 0.0524 & 56 & Extended high-ionization emission.\\
SDSS J084810.11+351534.3 & Sy 2 & 0.0573 & 72, 107 & KUG 0845+354. 2nd Sy 2 or cross-ionization,  AGN line ratios across extended region.\\
SDSS J104232.05+050241.9 & Sy 2 & 0.0272 & 98, 212 & NGC 3341. Extended emission into companion.\\
SDSS J110019.10+100250.7 & Sy 2 & 0.0361 & 133, 134 &  UGC 6081; 2 AGN, AGN-ionized cloud.\\
SDSS J120149.74-015327.5 & Sy 1 & 0.0907  & 139,142  & AGN is Sy 1.9, ionized tail opposite star-forming companion.\\
SDSS J121418.25+293146.7  & Sy 2 & 0.0632 & 60, 147  & Was 49; 2nd AGN or cross-ionization. \\
SDSS J134143.75+554025.5 & Sy 2 & 0.0251 & 98 & NGC 5278/9; 2 AGN, high [O III]/H$\beta$, He II in filaments.\\
SDSS J135429.05+132757.2 & Sy 2 & 0.0633 & 84, 186, 194 & Both galaxies have similar line ratios; [O III] in tail opposite companion.\\
    &          &                        \\
\multispan5 Other systems: \hfil \\
 & & & & \\
SDSS J075729.04+351105.9 & Sy 2 & 0.1117 & 99 & Double spatial/spectral line profiles.\\
SDSS J080004.05+232616.2 & Sy 2 & 0.0292 & 118 & Velocity structure in small emission region around AGN.\\
SDSS J083848.14+040734.0 & Sy 2 & 0.0476 & 96 & AGN emission extended 5\arcsec\ toward H II compn\\
SDSS J084135.08+010156.2 & Sy 2 & 0.1106 & 48 &  Two AGN, extended [O III] beyond each.\\
SDSS J084742.44+344504.4 & QSO & 0.0640 & 20 & PG 0844+349. Companion has resolved rotation curve, lacks [O III].\\
SDSS J090436.92+553602.9 & Sy 1 & 0.0372 & 7 & Emission around AGN extended by 15\arcsec\ , more toward absorption-ine companion.\\
SDSS J101043.36+061201.4 & Sy 1 & 0.0978 & 58 &  [O III] resolved over 8\arcsec\ . Weak emission in companion, no [O III].\\
SDSS J105030.47+232931.4 & Sy 1.8 & 0.0604 & 117 &  AGN unresolved. Star-forming companion.\\
SDSS J110157.90+101739.3 & Sy 1.5 & 0.0341  &138 & Extended rotation curve over 29\arcsec\ , star-forming line ratios.\\
SDSS J113240.25+525701.3 & Sy 2 & 0.0266 & 75 & Mkn 176. AGN emission extended to 6\arcsec\  opposite from 2 companions.\\
SDSS J113858.89+141253.2 & Sy 2 & 0.0805 & 10 &  Companion has resolved line emission, star-forming ratios.\\
SDSS J114252.83+325124.2 & Sy 2 & 0.0666 & 165 & Companion has extended emission, star-forming ratios.\\
SDSS J120041.39+314746.2 & Sy 2 & 0.1159 & 33 & AGN emission spans 14\arcsec\ , blends w/companion.\\
SDSS J121303.34+513854.9 & Sy 2 & 0.0849 & 118  & AGN unresolved. Star-forming companion emission resolved.\\
SDSS J121943.13+132659.9 & Sy 2 & 0.0647 & 50 & AGN em unresolved, companion has [N II] and Balmer absorption.\\
SDSS J124322.55+373858.0 & Sy 2 & 0.0859 & 120 & LINER-like AGN em to 8\arcsec\ opposite star-forming companion.\\
SDSS J130354.71-030631.8 & Sy 2 & 0.0778 & 166 & AGN emission not well resolved.\\
SDSS J134203.48+183901.5 & Sy 1.8 & 0.0845 & 102  & AGN em (low-ion NLR) to 8\arcsec opposite star-forming companion.\\
SDSS J134736.39+173404.6 & Sy 2 & 0.0447 & 108 & AGN-ionized knot at 18\arcsec opposite star-forming companion.\\
SDSS J135255.67+252859.6 & Sy 1.5 & 0.0636 & 176  & Strong lines like starbursts, but He II/H$\beta \approx 0.2$ and broad H$\alpha$.\\
SDSS J141447.15-000013.1 & Sy 2 & 0.0475 & 65 & Sy 2 plus transition nucleus w/[O III]/H$\alpha \approx 2$. Emission unresolved.\\
Kaz 199    & --- & 0.0155 & 5, 54 & NGC 6636. [O I]/H$\alpha \approx  0.03$, may not be AGN\\
\hline
\end{tabular}
\label{tbl-spectra}
\end{table*}

\begin{figure*} 
\includegraphics[width=170.mm,angle=0]{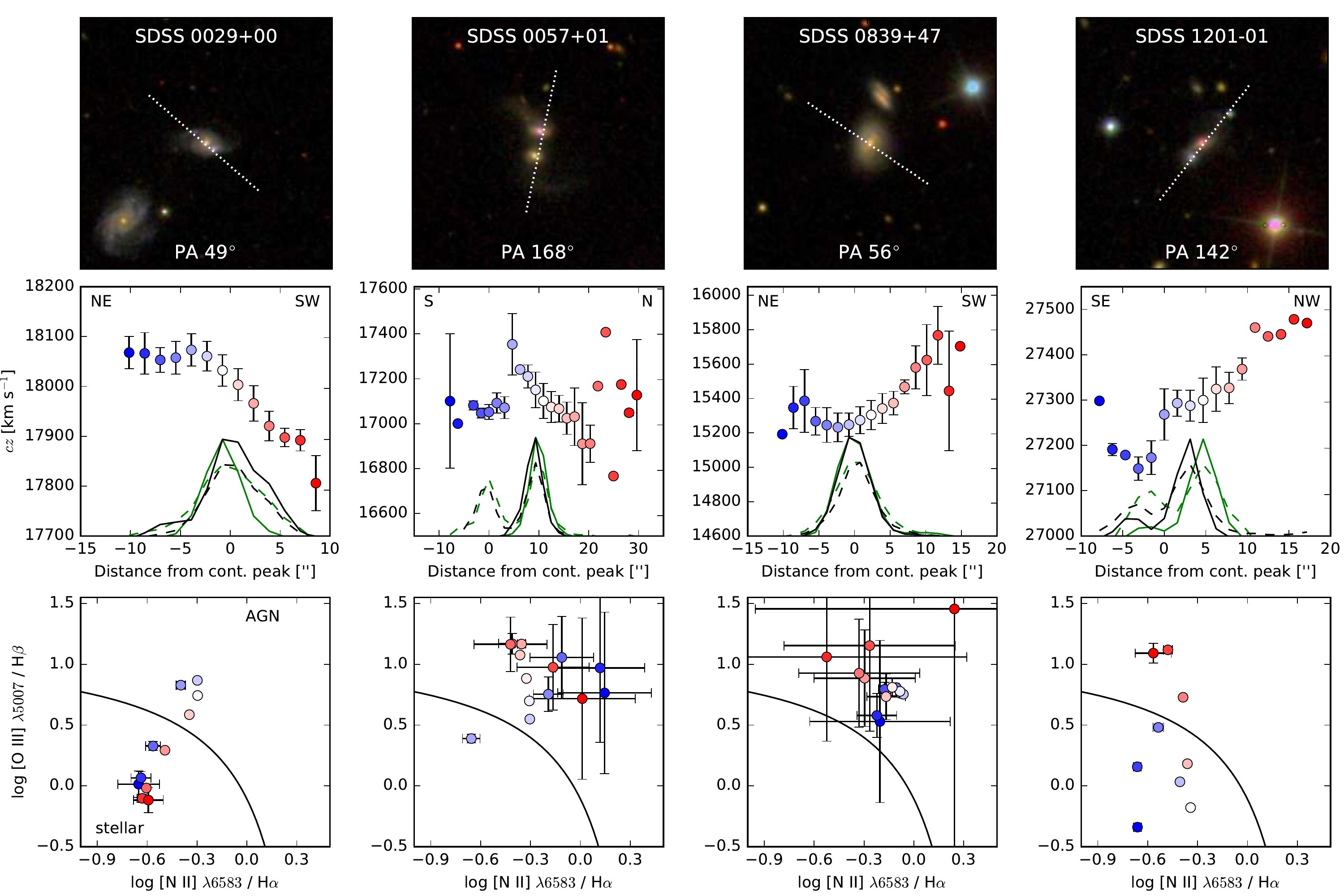} 
\caption{Results of long-slit spectroscopy of candidate AGN cross-ionization systems. For each object, the top panel shows the 
slit position on an SDSS DR12 $gri$ composite image spanning $100 \times 100$\arcsec. The middle panel shows the heliocentric radial-velocity profile along the slit, with error bars based on agreement among multiple lines measured at each position,
and colour coding of points to indicate position. Below the velocity profiles are intensity traces of
H$\alpha$ (black) and [O III] (green, full lines), with the adjacent continuum (dashed in matching colour) for each, on arbitrary linear 
vertical scales, to show the alignment between intensity and velocity features. The bottom
panels show the strong-line BPT classification diagram with the starburst/AGN dividing curve from
\citet{Kewley2006}; point colors here show where along the slit AGN-ionized regions occur
in comparison with the middle panels. The H$\beta$ flux was estimated from H$\alpha$
and an intrinsic Balmer decrement 2.85; additional reddening correction would move data points
upward and toward (or further into) the AGN-ionized region above the curves.
Regions appear in this panel only if H$\alpha$, [O III] and [N II]
are all detected.}
 \label{fig-longslit}
\end{figure*} 


\begin{figure*} 
\includegraphics[width=170.mm,angle=0]{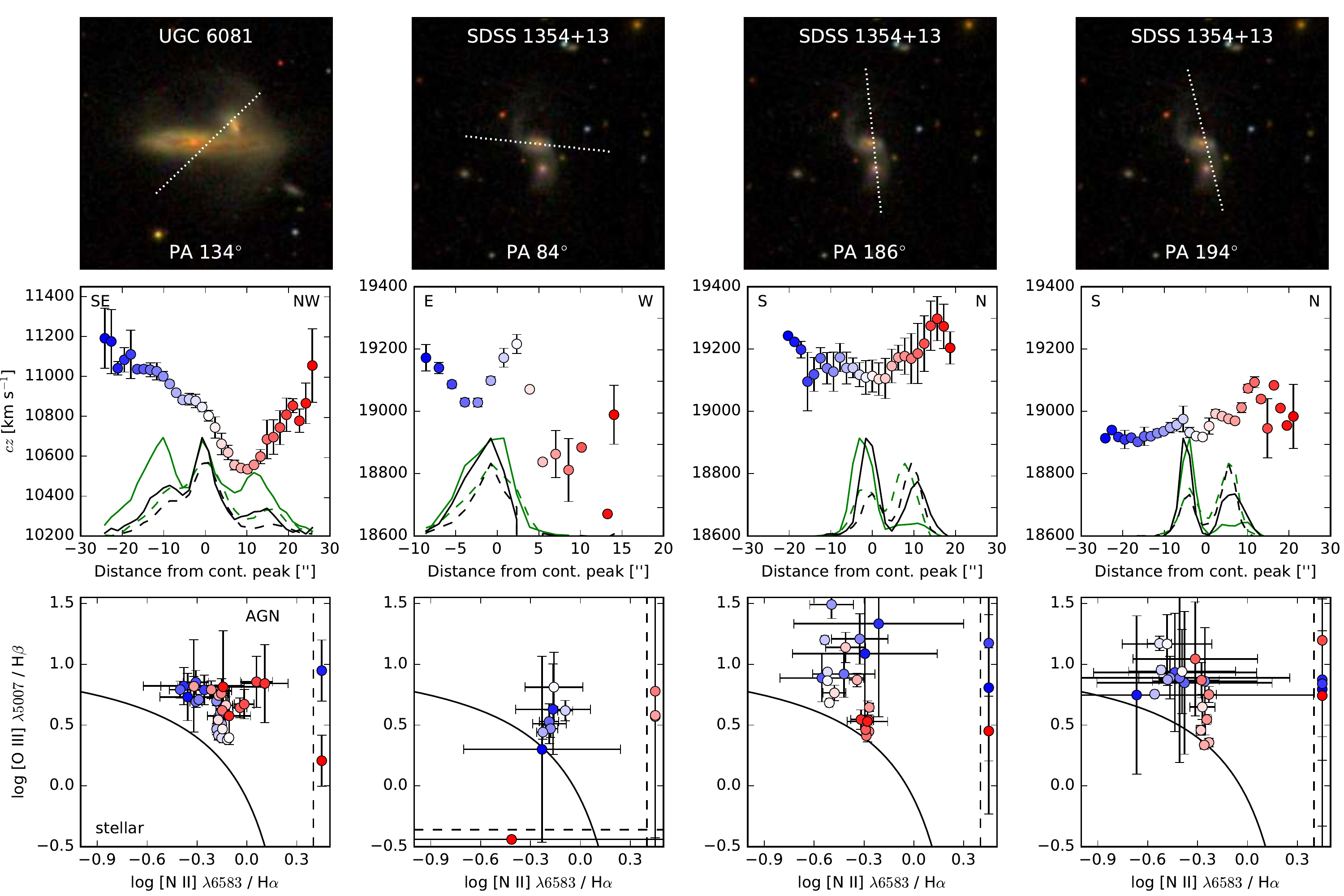} 
\contcaption{} 
\end{figure*} 


\begin{figure*} 
\includegraphics[width=170.mm,angle=0]{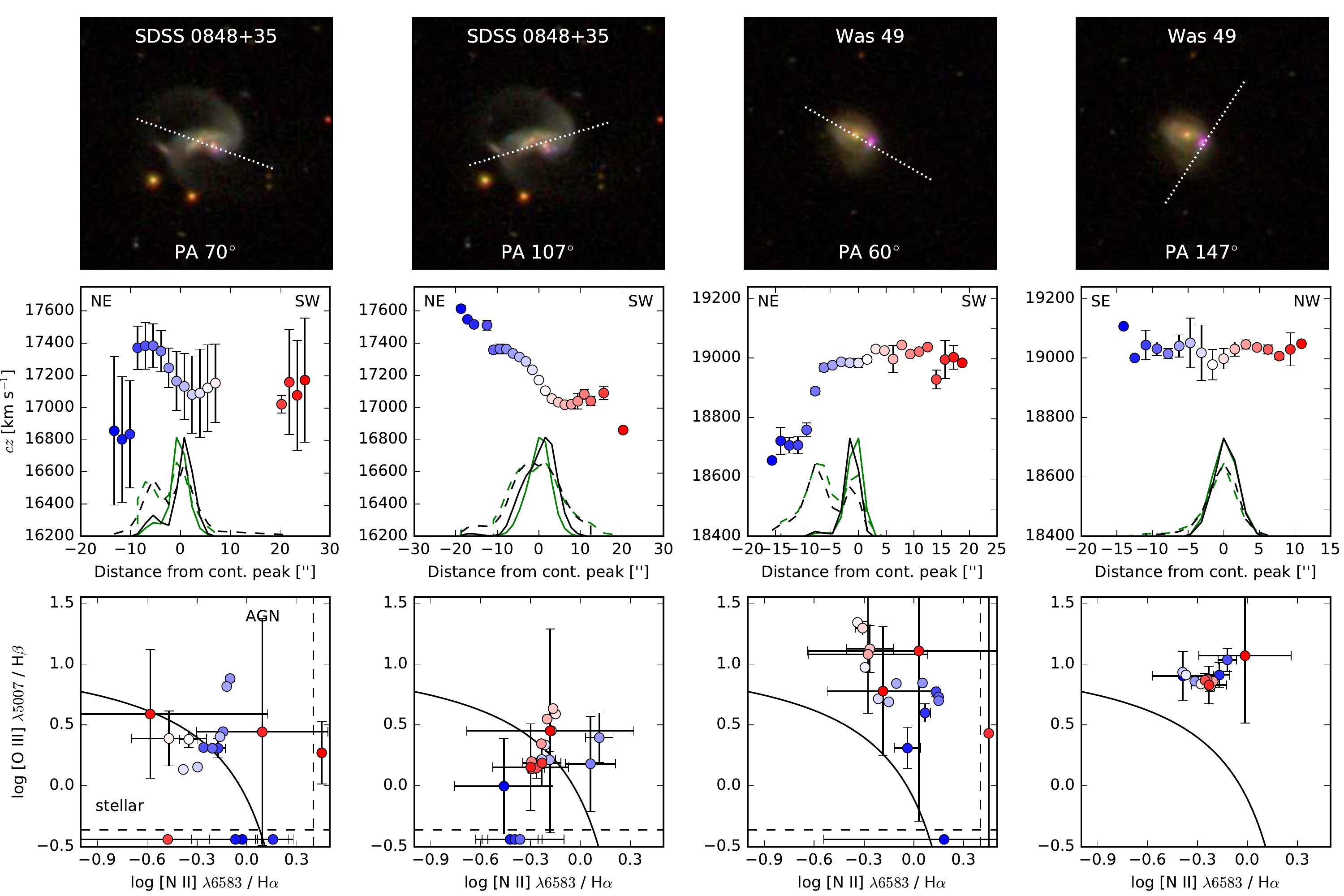} 
\contcaption{} 
\end{figure*} 


\begin{figure*} 
\includegraphics[width=170.mm,angle=0]{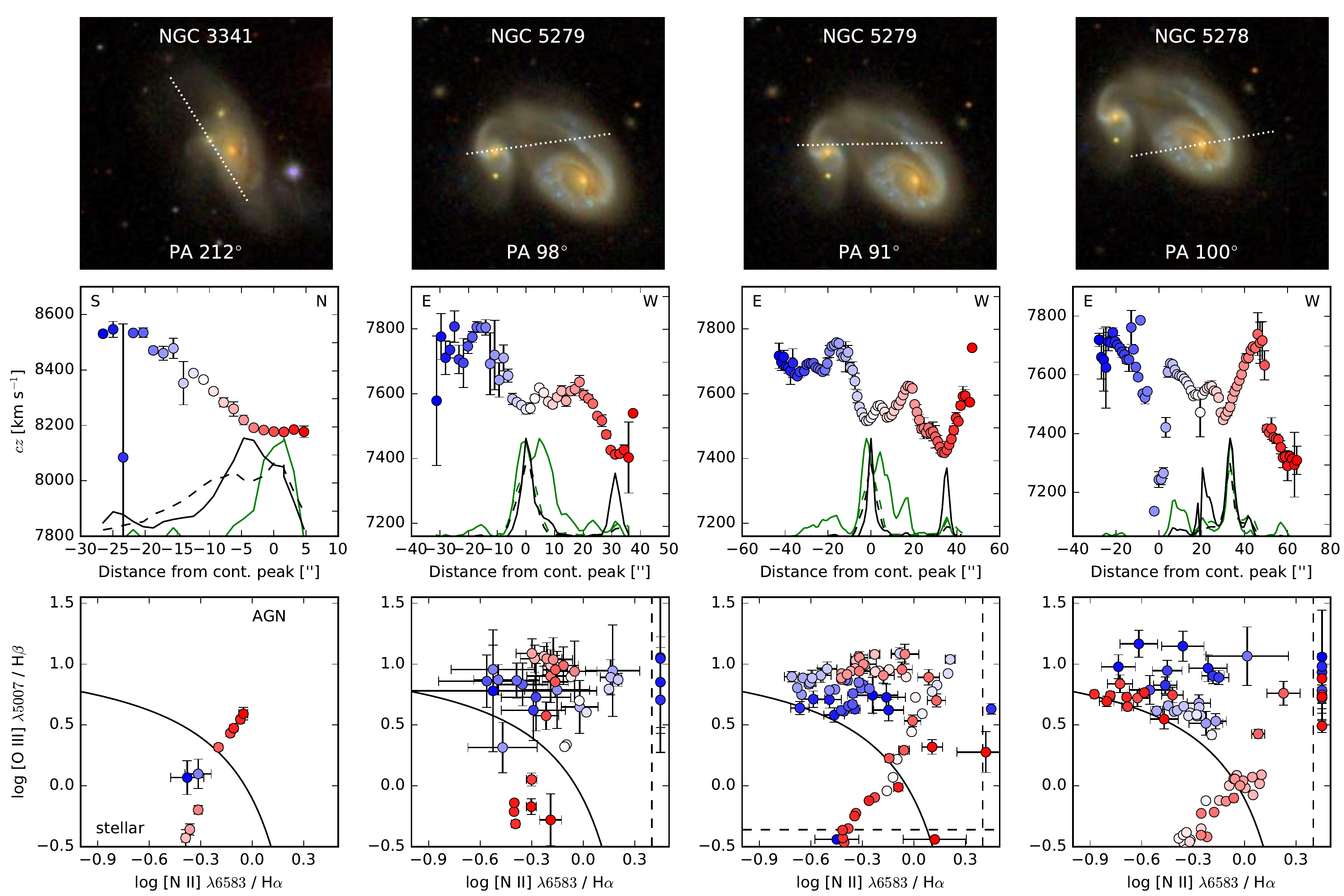} 
\contcaption{} 
\end{figure*} 

\subsection{Observations: narrowband imaging}

Our spectroscopy is supplemented by narrowband imaging for 8 systems, summarized in Table \ref{tbl-narrowband}.
For seven of the objects with the highest priority for spectroscopic observations in the redshift range $z = 0.008 - 0.027$, where [O III]
falls in the band of an available filter, we obtained images in
redshifted [O III] and the V-band continuum using the SARA 1-m remote telescope at Kitt Peak,
Arizona and the 1-m Jacobus Kapteyn telescope at La Palma, now operated by SARA \citep{SARA}. Narrowband exposures in a filter
of 100-\AA\  FWHM were 1-2 hours total, with 15-30 minutes in V.
Among these, we recover the well-known ionization cones in NGC 2992 
(\citealt{WehrleMorris}, \citealt{Allen}, \citealt{Veilleux})
perpendicular to its disc plane (but not oriented toward either its tidal tail or companion). We find only nuclear emission from Mkn 176.
NGC 3786/8 and UGC 3995 A/B show [O III] emission only from the nuclei and knots in the spiral arms, as is typical for spirals.
On this basis, we did not obtain new long-slit spectroscopy of Mkn 176, NGC 2992, NGC 3788, or UGC 3995.

\begin{figure*} 
\includegraphics[width=130.mm,angle=270]{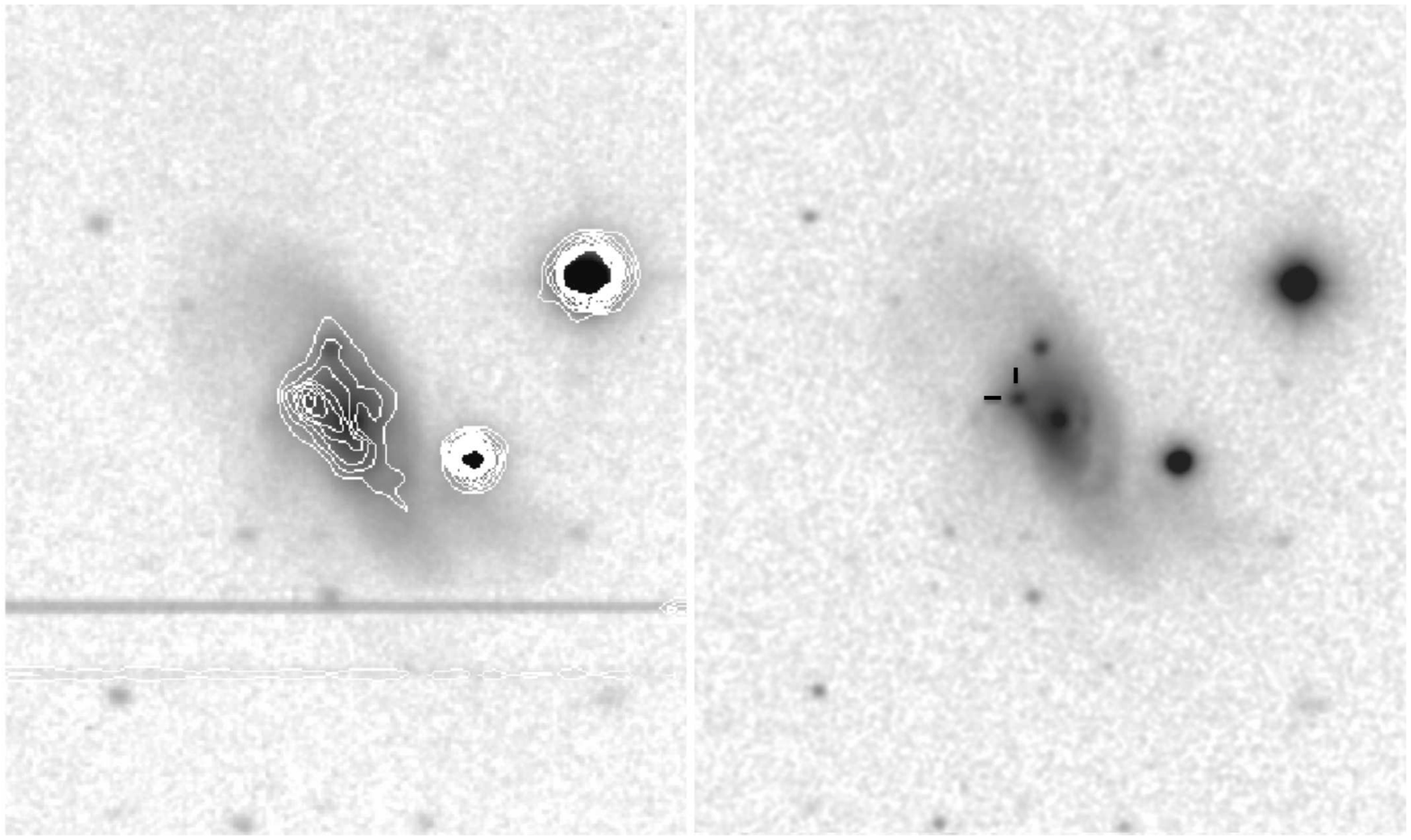} 
\caption{[O III] structure (linearly spaced contours) superimposed on the matched-resolution $V$-band continuum of NGC 3341, with 
a higher-resolution
$V$ image from the JKT also shown for comparison of the inner structure. The Seyfert nucleus in this system 
is in the small companion galaxy to the east (indicated by tick marks in 
the right-hand panel); 
the [O III] structure may represent an ionization cone comprised of material mainly in the disc of the large galaxy. Stellar residuals result from colour differences
between foreground stars and the galaxies; two bright-star charge trails are seen on the southern side.
The area shown is $167 \times 201${\arcsec}, with north at the top. The intensity mapping in $V$ uses a logarithmic scale
offset from zero to avoid amplifying noise around the sky level.}
 \label{fig-ngc3341o3}
\end{figure*}

NGC 3341 shows a potential morphological signature of cross-ionization. The AGN here is in a small galaxy projected against the disc of a much larger spiral.
In [O III], there is a triangular region of emission crossing the large disc at a skew angle with its apex at the AGN (Fig. \ref{fig-ngc3341o3}), 
which we tentatively interpret as an ionization cone of gas in the large galaxy. The slit location for spectroscopy was selected to
sample this region, rather than crossing the nucleus of the large companion galaxy.

Motivated by extended [O III] structure in our SARA images of Kaz 199, we obtained more detailed images with the KPNO 2.1-m telescope. [O III] was 
isolated with a filter of 174 \AA\  FWHM centered at 5142 \AA\ , and H$\alpha$ with a filter of 68 \AA\  FWHM centered at 6653. V and R were used 
for continuum subtraction.

For NGC 5278/9, we also analyze the H$\alpha$ and continuum images from the KPNO 2.1-m telescope described by \cite{KKHHR87},
which guided our slit placement on noting that this pair was included in the AGN/companion sample. Our spectroscopic data,
designed to encompass three of these near-radial filaments,
help interpret the nature of the emission filaments mentioned by \cite{KKHH89}. Additional images in
[O III] were obtained with the SARA system.

Emission-line mapping of UGC 6081 was carried out on Apr 16 2018 at the 2.5-m telescope of the Sternberg Astronomical Institute (SAI) of
the Caucasus Mountain Observatory (CMO) on Mt. Shatdzhatmaz  \citep{Kornilov2014}  with the the Mapper of Narrow Galaxy Lines (MaNGaL). MaNGaL (Moiseev \& Perepelitsyn, in prep.) is   a tunable--filter imager based on a scanning Fabry-Perot interferometer with low interference 
order\footnote{see https://www.sao.ru/Doc-en/Events/2017/Moiseev/moiseev\_eng.html}. The peak of filter transmission (FWHM$\approx15$\AA) was centered on the wavelength corresponding to the redshifted [OIII]$\lambda5007$ and H$\alpha$ emission lines. Continuum images were consecutively exposed at wavelengths shifted by $\sim50$ \AA\  and subtracted to produce net emission-line images. 

\begin{table*}
 \centering
  \caption{Narrowband imaging}
  \begin{tabular}{@{}lccccl@{}}
  \hline
 Object	&	$z$ & 	Bands &		Exposure [min] & 	Telescope & 	UT date\\
 \hline
Kaz 199  & 	0.0155	& V/510 		& 20/30	& SARA-KP 	 &  25 Feb 2012\\
Kaz 199  & 0.0155  	& V/514  		& 10/30    & KPNO 2.1m & 13 Apr 2013\\
Kaz 199 &		0.0155	& R/6653		& 10/45	& KPNO 2.1m  & 29 May 2012\\
Mkn 176  & 	0.0274	& V/510 		& 20/60	& SARA-KP 	 & 12 Mar 2013\\
NGC 2992  &	0.0077	& V/510		& 15/120	& SARA-KP 	  & 2 Feb 2013   \\
NGC 3341  &	0.0273	& V/510 		& 30/120	& SARA-KP 	  & 2 Feb 2013   \\
NGC 3341 &	0.0273	& V/510 		& 20/60	& SARA-KP 	  & 29 Dec 2014\\   
NGC 3341  & 0.0273   & B/V   &    30/30  & SARA-JKT & 4 Feb 2017\\
NGC 3786  &	0.0089 	& V/510		& 20/60	& SARA-KP	  & 4 Apr 13   \\
NGC 5278/9 &	0.0251 	& 6000/6693	& 10/20    & KPNO 2.1m    & 18 May 1985\\
NGC 5278/9 & 0.0251    & V/510            & 10/60     & SARA-KP      & 27 Mar 2015 \\
UGC 3995  &	0.0158	& V/510		& 20/60	& SARA-KP	  & 31 Dec 2013\\ 
UGC 6081 & 0.0361 & H$\alpha$/continuum  & 33/33 & CMO 2.5m & 16 Apr 2018 \\ 
UGC 6081 & 0.0361 & [O III]/continuum  & 33/33 & SAI 2.5m & 16 Apr 2018 \\ 
\hline
\end{tabular}
\label{tbl-narrowband}
\end{table*}

For galaxies or regions with large enough equivalent width in the [O III] lines, we can produce [O III] images from linear combinations of the
SDSS survey images (as shown by \citealt{mnras2012}, and addressed systematically by \citealt{Sun2018}). We do this for  Was 49 (Fig. \ref{fig-was49o3}), revealing extensions from the AGN in opposite directions, possibly including gas in the large companion disc.

\section{Cross-ionization (and missing-ionization) candidate systems}

Several of the systems in Table {\ref{tbl-spectra} show evidence of cross-ionization. Others
show no such 
evidence despite high {\it a priori} odds of seeing it from the system properties. We discuss
some of their individual properties here.

{\bf SDSS 0029+00 (UM 246)}: A second spectrum of the companion galaxy to the south shows a potential AGN signature on the
northern side of its disk. The spectrum of the AGN itself shown in Fig. \ref{fig-longslit} includes tidal arms, whose emission-line
ratios indicate dominant role for local star formation.

{\bf SDSS 0057+01 (UM 293)}: AGN-ionized line ratios extend over most of the region sampled in both galaxies. 
The outermost parts at both ends of the slit are kinematically distinct from the rotational signature
of the AGN host itself.

{\bf SDSS 0839+47}: our spectrum across the AGN host shows that the AGN ionizes gas across the
galaxy, with increasing ionization in the outer parts. The SDSS spectrum of the companion galaxy 
shows line ratios indicating star formation, albeit with [O I] rather strong at about 0.07 $\times$
H$\alpha$, broadly consistent with mixed ionization sources. This system was
flagged by \cite{Comerford14} for showing velocity asymmetries in both Balmer and
forbidden emission lines.

{\bf SDSS 1201+01}: the companion galaxy shows line ratios indicating ionization by hot stars. Gas in the tidal
tail on the opposite side of the Sy 1.9 AGN is photoionized by the AGN. 

{\bf UGC 6081}: This pair has two heavily reddened Sy 2 nuclei and shows AGN ionization along
an extent of 58" (41 kpc). The extended emission might fall in the same category of
radial extent $> 10$ kpc
as the ``Voorwerpjes"  \citep{mnras2012}, depending  on which of the AGN
lights it up. We do not have the three-dimensional information to tell which is
responsible (if indeed one dominates in the tidal debris seen at large radii). We do
have the information for a rough ionization-balance calculation as done above for
NGC 5278/9. These nuclei are more luminous and much more heavily reddened than
in NGC 5278/9; applying a single foreground-screen reddening correction based on the nuclear Balmer decrements suggests that
the NW AGN, in the smaller galaxy (UGC 6081 NED01), is twice as luminous in 
[O III]. Our estimates (Table \ref{tbl-balance},
listing dereddened values in this system)
show that the energy balance makes it plausible for the SE AGN (in UGC 6081 NED02) to 
power the 
most distant gas we observe spectroscopically along the slit to its southeast, although
whether this works in detail depends critically on the fine structure of the gas (since
the peak surface brightness we observe places the strongest limits on the incident flux,
and is strongly dependent on image quality). Much of the reddening toward each nucleus
may take place close to our line of sight rather than as surrounding ``cocoons" covering large
solid angles; the WISE data suggest that in neither one does the MIR component of the SED dominate
over ionizing radiation reaching surrounding gas. Using our conservative fine-structure correction (a factor 3
as noted above), the case for AGN fading in this system is weak (energy shortfall 1.7-2.0 times
depending on which AGN is responsible, and no shortfall if both contribute to the ionization of the extended cloud).

The tunable-filter data are compared in Fig. \ref{fig-ugc6081}. They reveal rich structure around both galaxies, with
emission extending 25\arcsec (18 kpc) from both AGN. The emission regions occur within the tidal debris and loop
seen in the starlight continuum, fitting with a common pattern of kinematically quiescent EELRs as photoionized
tidal tails. Within the host galaxy discs, H$\alpha$ dominates over [O III] emission, which is reversed outside
the galaxy discs; there are distinct bright knots of higher [O III]/H$\alpha$ within the extended clouds. The structure
we see does not make clear to what extent each AGN contributes to the overall ionization, except that it occurs
on only one side of the NW AGN and more symmetrically about the SE one.


\begin{figure*} 
\includegraphics[width=130.mm,angle=0]{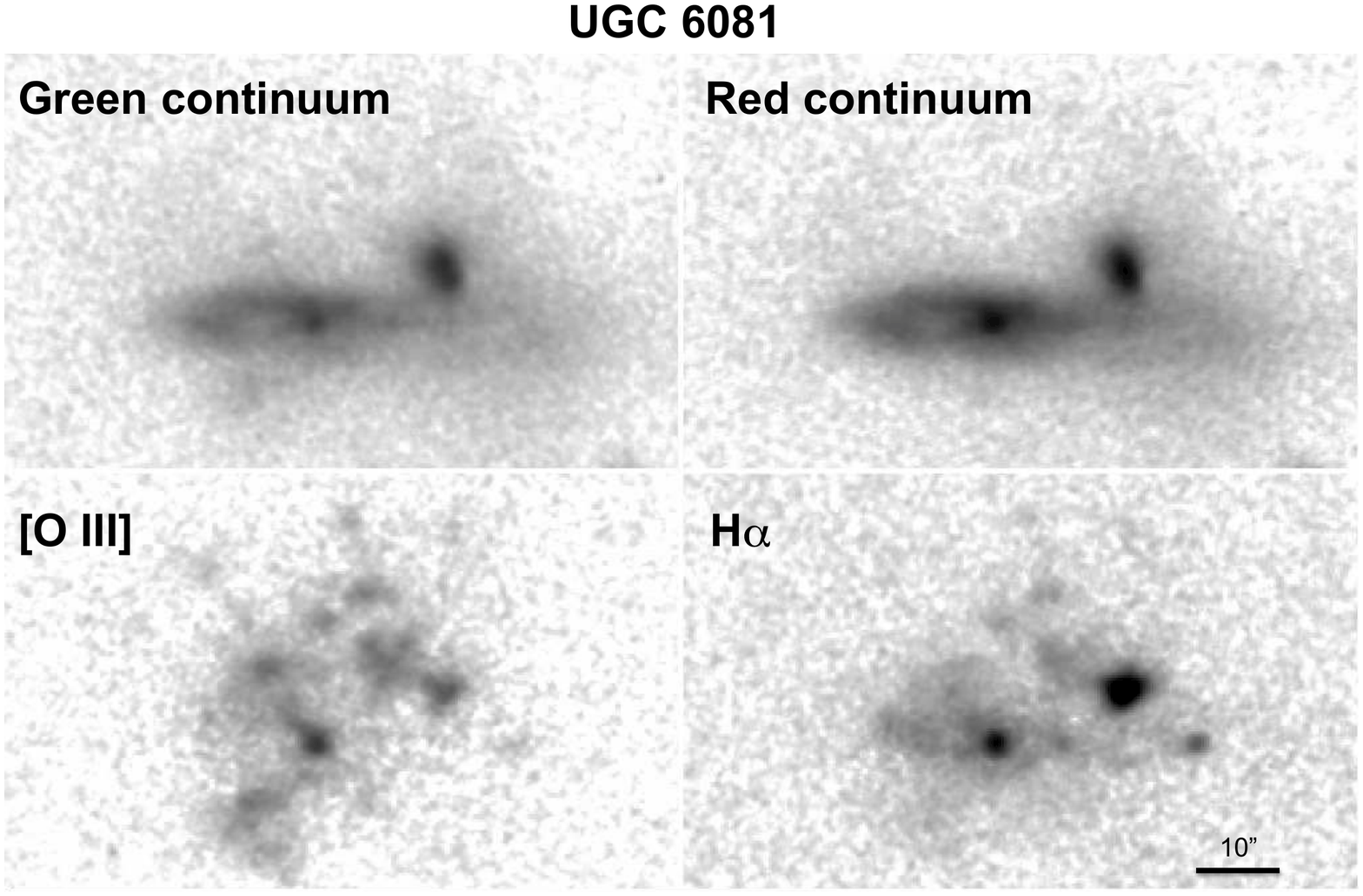} 
\caption{The SAI 2.5m tunable-filter data for UGC 6081. Continuum images are offset by 50 \AA\
from the redshifted line wavelengths, and bottom images show net line emission after continuum subtraction. North is at the top
and east to the left; each panel spans $50 \times 82$\arcsec. The bright nucleus of the southeastern disc is bright in all these bands and 
gives a convenient reference point for comparison.
Emission regions appear projected as far as 25\arcsec\ 
from each nucleus, so that even if the nearer one is solely responsible for the photoionization at these locations, the radial extent 
from the ionizing source is at least
18 kpc. The intensity scales are similar to Sloan image products, changing gradually from linear at low intensities to logarithmic at bright levels.}
 \label{fig-ugc6081}
\end{figure*} 

{\bf SDSS 1354+13}: both galaxies show similar line ratios, marking a possible dual AGN system
flagged by \cite{Liu} from the SDSS spectrum, and with a velocity asymmetry in the
SDSS spectrum noted by \cite{Comerford14}. AGN-ionized
gas is more extended than the starlight continuum, notably in the tidal tail to the south, where
it is detected out to a radius of 22" (27 kpc) along two adjacent slit positions. \cite{Comerford17} recently
presented a suite of spectroscopic data showing both an EELR and a spatially resolved outflow,
with implied timescales suggesting recurrent episodes of nuclear activity in which the AGN has turned ``off" and ``on" again
within $10^5$ years. Their EELR detection south of the nucleus
overlaps with our data along the N--S slit position; our detection extends slightly farther from the AGN. Both their data set and ours
indicate that the EELR is part of a tidal tail. 

{\bf SDSS 0848+35}: there is an isolated region of AGN-ionized gas at a projected radius 21" (23 kpc).

{\bf Was 49}: Identified in this context by \cite{Moran}, this may remain the best low-redshift
cross-ionization candidate. Detailed study of the AGN by \cite{Secrest} stresses its high luminosity and implied black-hole mass relative to the very modest stellar mass of the host galaxy.
The [O III] structure around the AGN is elongated
perpendicular to the line between the galaxies (Fig. \ref{fig-was49o3}); the
images in \cite{Secrest} show that part of this emission is in a single bright discrete source just northwest
of the AGN.
All the locations along our spectroscopic slits in this pair are AGN-ionized except the
nucleus of the large companion galaxy; the AGN emission is largely continuous with the
portion of the large galaxy's rotation curve sampled in our data. Integral-field spectroscopy with 
high angular resolution could show whether the gas seen surrounding the AGN on arcsecond
scales kinematically matches either its own radial velocity or the rotation pattern of the larger
disc. In the latter case, the ionization pattern in the disc might reflect its intersection with the pattern of
escaping ionizing radiation.

\begin{figure} 
\includegraphics[width=65.mm,angle=270]{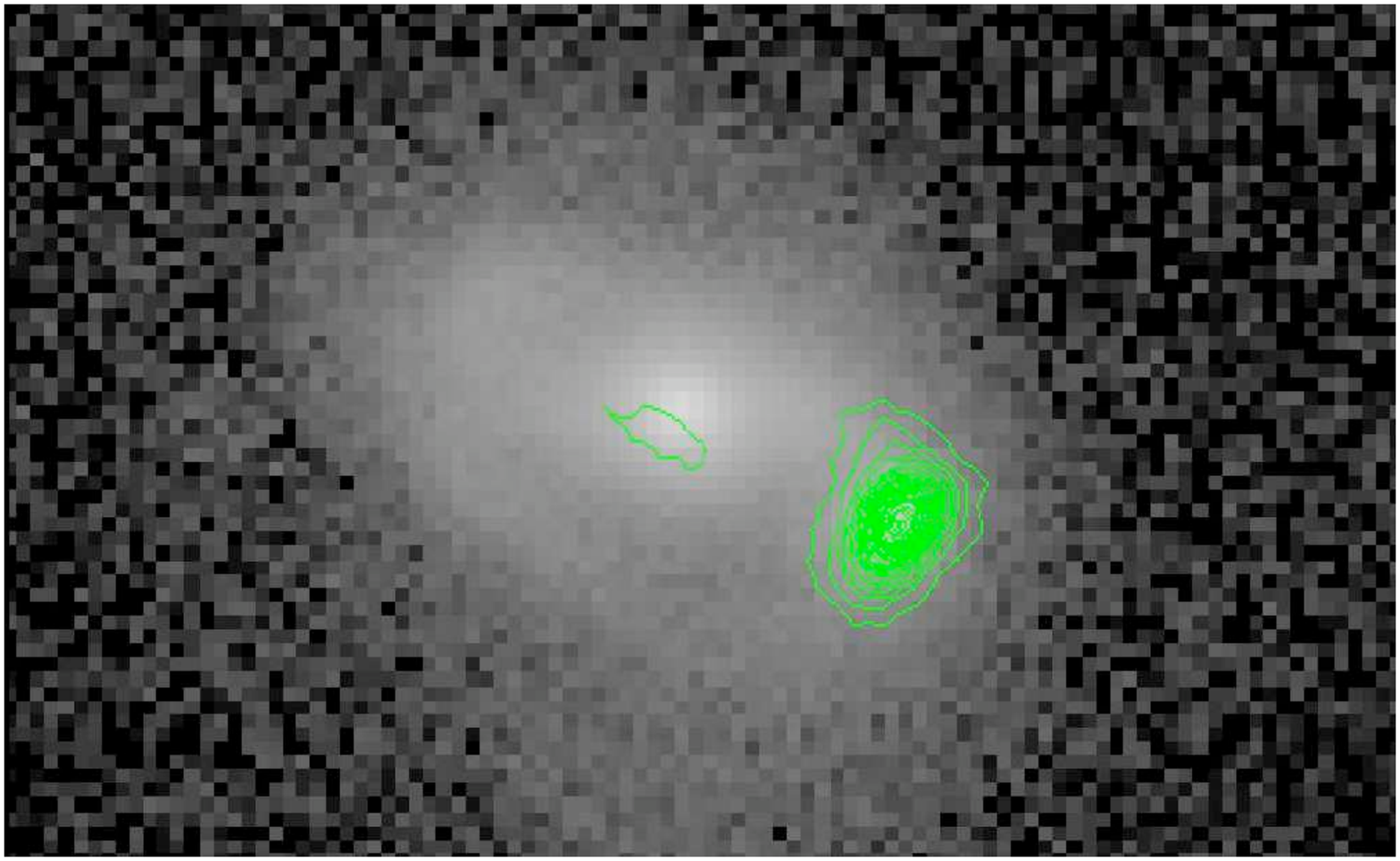} 
\caption{[O III] structure (contours) superimposed on the $r$-band image of Was 49, from
SDSS data using the difference between $g$ and $r$ data to trace line emission. The Seyfert nucleus in this system is in the small companion galaxy to the west, superimposed on the large companion disc and centered in the contours; 
the [O III] structure may represent an ionization cone
The area shown is $26 \times 45${\arcsec}, with north at the top.}
 \label{fig-was49o3}
\end{figure}

{\bf NGC 3341}: This pair has nearly the ideal geometry from Fig. \ref{fig-schematic}, with
an AGN hosted in a compact galaxy close to a much larger spiral \citep{Barth08}. In
a multiwavelength study,  \cite{Bianchi13} set very low limits on AGN in the other components of this system. Our emission-line imaging
(Fig. \ref{fig-ngc3341o3}) shows a possible ionization cone in [O III] southward from the AGN across the larger disc. However, the spectroscopic data show AGN ionization dominant only close to the AGN, so this structure may be an accident of the spiral pattern
and location of star-forming regions. On the other hand, the spectrum shows unusually strong [O I] emission for pure stellar photoionization 
across much of this region, with [O I] $\lambda 6300$/H$\alpha \approx 0.03$. More detailed data and modeling 
would be needed to tell whether this is low-metallicity gas ionized by an AGN, a composite of both ionization
sources blended along the slit, or something yet more  complex.

{\bf NGC 5278/9}: This pair has two low-luminosity AGN, near the Sy 2/LINER boundary in
line-ratio diagrams depending on aperture size. The system shows extensive 
filaments or plumes in line emission, some almost radial (\citealt{KKHH89}, Fig. \ref{fig-ngc5279}).
Our spectroscopy in the BPT diagram shows AGN-ionized gas across the pair, including one of these plumes
to the east. This may suggest that either the AGN are heavily obscured along our line
of sight or have faded over the light-travel time from the nucleus to the extended emission regions. Because gas in each disc may 
be largely shielded from its own AGN, this pair is a potential instance of cross-ionization.

In this system, we have the data to estimate both the nuclear luminosities and the
ionizing-flux requirement at various points in the extended emission, to see whether
the nuclear output has changed dramatically over the relevant light-travel time, and
how important obscuration of the AGN is along our line of sight. Since these nuclei
were not observed spectroscopically in the SDSS, we use the optical emission-line data
within 4\farcs 7 apertures reported by \cite{KKHH85}. For mid-infrared nuclear
properties, we take the PSF-fitting WISE fluxes \citep{Wright2010}, using the ``compromise"
spectral slope conversions from \cite{cutri2012} and bearing in mind the potential non-nuclear contribution 
within the WISE PSF. Conversion from H$\alpha$ surface 
brightness in extended regions to required luminosity in the ionizing continuum followed
\cite{mnras2012}. To be conservative, we take the distance from extended cloud regions to 
each nucleus to be in the plane of the sky, minimizing the ionizing-luminosity requirements.
The quantities involved in this comparison are given in Table \ref{tbl-balance}.

This kind of estimate gives only a lower bound to the required ionizing luminosity of the AGN at the relevant time,
both because a given parcel of gas is unlikely to be optically thick at the Lyman edge, and because we measure
spatially averaged surface brightness in structures that are likely to be very patchy. For example,  
\cite{Gagne} and \cite{Keel2017b}, working at a linear resolution of $\approx 150$ pc from HST imaging of the ``Teacup" AGN,
find that this calculation for the brightest regions underestimates the ionizing continuum by about a factor 3 over a wide range in radius.
For NGC 5278/9, we average over a yet larger region, roughly 2 kpc in size. Lacking more detailed information on the
fine structure in NGC 5278/9, we consider that the multiplication factor will be at least 3.
However, with the evidence from He II emission that the extended features from both galaxies are AGN-ionized,
a fading scenario would be plausible only if one of the AGN dominates the ionizing output, since it would be unlikely for both 
AGN to have behavior synchronized at the $10^4$-year level, much faster than dynamical timescales associated with the galaxy encounter.

The deep BTA spectra, centered on each nucleus and aligned along the near-radial filaments, not only extend the radial
range over which we can locate feature on the BPT diagram, but add key new information - in three of  these filaments,
extending outside the detected stellar discs, the He II/H$\beta$ ratio is so high that it must arise from
photoionization by an AGN continuum (Table \ref{tbl-NGC5278}). The regions in NGC 5278 fall near the 
boundary between stellar and AGN photoionization in the BPT diagram, illustrating the effects of low metallicity as this primarily
reduces the strength of [N II].

\begin{table*}
 \centering
  \caption{Recombination-balance analysis}
  \begin{tabular}{@{}lcccc@{}}
  \hline
 Quantity&	 NGC 5278 & 	NGC 5279 &	UGC 6081 SE & UGC 6081 NW \\
 \hline

H$\alpha$ flux, nucleus (erg cm$^{-2}$ s$^{-1}$)  							& $2.17 \times 10^{-15}$	& $2.20 \times 10^{-15}$		& $5.09 \times 10^{-14}$	& $2.86 \times 10^{-13}$\\
\ [O III] $\lambda 5007$ flux, nucleus (erg cm$^{-2}$ s$^{-1}$)	& $9.7 \times 10^{-15}$	& $1.2 \times 10^{-14}$	& $1.6 \times 10^{-14}$	& $1.44 \times 10^{-13}$\\
L(ion) from [ O III]	(erg s$^{-1}$)					& $4.9 \times 10^{42}$		& $6.2 \times 10^{42}$	& $9.4 \times 10^{43}$	& $1.9 \times 10^{44}$\\
Q(ion) from [ O III]	(photons s$^{-1}$)			& $1.1 \times 10^{53}$		& $1.4 \times 10^{53}$	& $2.2 \times 10^{54}$	& $4.4 \times 10^{54}$\\
F(MIR) (8-25 $\mu $m) (erg cm$^{-2}$ s$^{-1}$)	& $1.5 \times 10^{-11}$		& $8.5 \times 10^{-12}$ & $2.2 \times 10^{-11}$& $3.0 \times 10^{-11}$\\
L(MIR)	(erg s$^{-1}$)					& $1.5 \times 10^{43}$		& $8.8 \times 10^{42}$ & $6.3 \times 10^{43}$ & $8.6 \times 10^{43}$\\
Max L(ion)	(erg s$^{-1}$)				& $2.0 \times 10^{43}$		& $1.5 \times 10^{43}$ & $1.6 \times 10^{44}$ & $2.8 \times 10^{44}$\\
Extended region needs: \\
L(ion)	(erg s$^{-1}$)					& $>4.2 \times 10^{43}$	& $>1.4 \times 10^{43}$	& $>9.4 \times 10^{43}$	& $>1.9 \times 10^{44}$\\
$3 \times$ structure correction (erg s$^{-1}$)			& $1.3 \times 10^{45}$		& $4.2 \times 10^{44}$		& $2.8 \times 10^{44}$	& $5.7 \times 10^{44}$\\			
\hline
\end{tabular}
\label{tbl-balance}
\end{table*}

\begin{table*}
 \centering
  \caption{Emission -line ratios in extended filaments of NGC 5278/9}
  \begin{tabular}{@{}lccc@{}}
  \hline
 Ratio & NGC 5278, PA 100,  22-29\arcsec\ E & NGC 5278, PA100, 20-27\arcsec\ W & NGC 5278, PA100, 19-28\arcsec\ E  \\
 \hline

{} He II $\lambda 4686$/H$\beta$ & $0.24 \pm 0.05$ & $0.54 \pm 0.15$ &  $0.043 \pm 0.08$ \\
{} [O III] $\lambda 5007$/H$\beta$ &  $ 4.69 \pm 0.25$ &  $6.11 \pm 0.75$ & $10.69 \pm 0.75$ \\
{} [O III] $\lambda 5007$/H$\alpha$ & $1.44 \pm 0.05$ & $1.87 \pm 0.08$ & $3.24 \pm 0.07$  \\
{} [N II] $\lambda 6583$/H$\alpha$ & $0.47 \pm 0.05$ & $0.18 \pm 0.03$ & $0.43 \pm 0.03$ \\
\hline
\end{tabular}
\label{tbl-NGC5278}
\end{table*}

\begin{figure} 
\includegraphics[width=100.mm,angle=0]{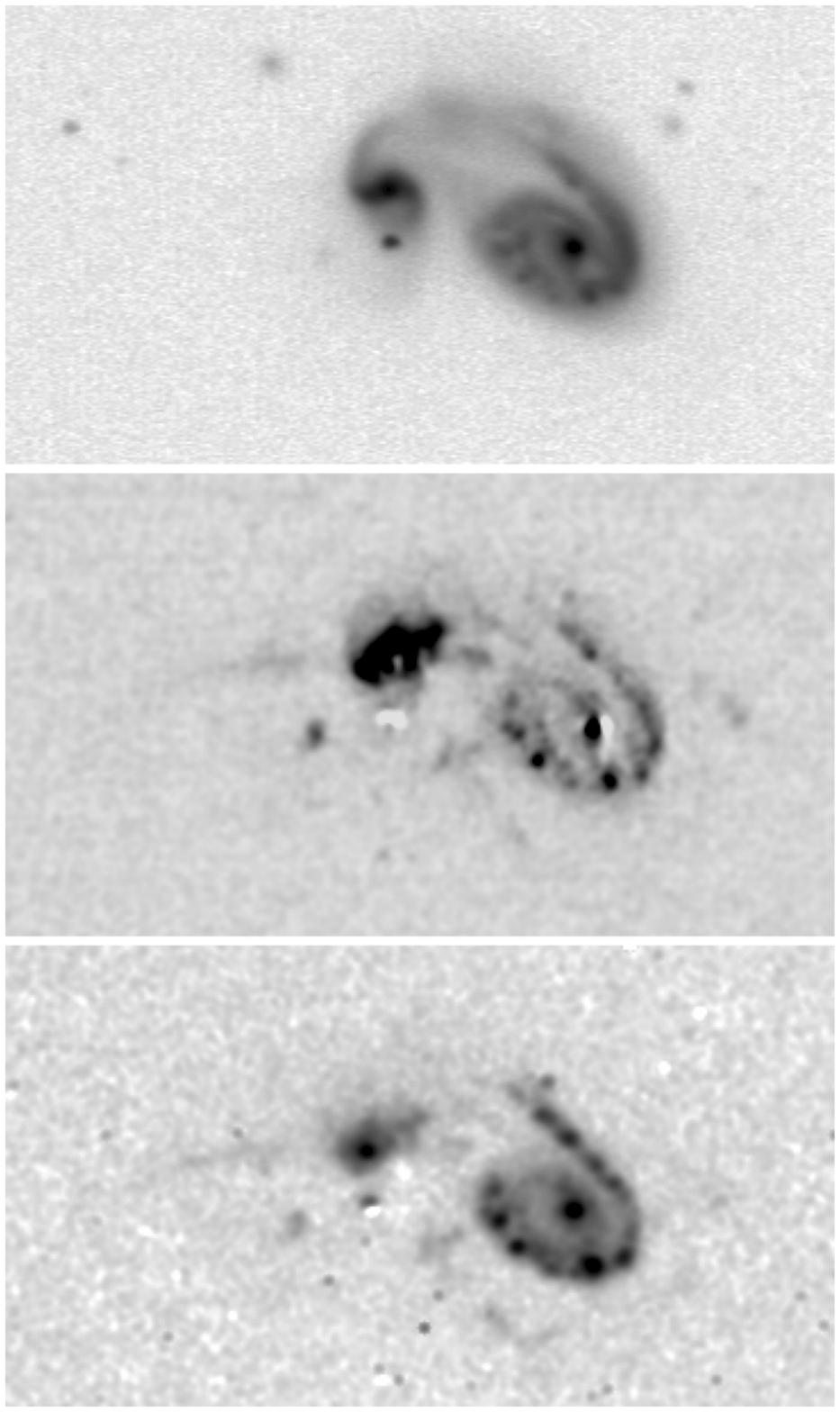} 
\caption{The NGC 5278/9 system in the $V$ band (top), continuum-subtracted [O III] (middle), and continuum-subtracted H$\alpha$+[N II] (bottom). The field shown is 141" across with north at the top. PSF mismatches between on-band and continuum images give artifacts at the 
locations of bright stars and the galaxy nuclei. At least three
partly radial emission features are present without significant continuum counterparts; our 
spectrograph slits crossed the nuclei both of NGC 5279, included much of the emission feature
extending to its east, and NGC 5278, covering the two oppositely-directed emission filaments to the ESE and WNW. The intensity 
scales are logarithmic starting at levels slightly offset from zero.}
 \label{fig-ngc5279}
\end{figure}

Among galaxies that our spectroscopic suggested are pairs of AGN rather than instances of cross-ionization,
HST images recently presented by \cite{SB2018} indicate that SDSS J084135.04+010156.3 may instead show cross-ionization.
Their [O III] and H$\alpha$ images show that the outer part of the 10-kpc ionization cone from the AGN coincides
with the companion galaxy in structure and brightness enhancement, suggesting that at the limited spatial resolution
of our data, that emission masquerades as a second AGN.

{\bf NGC 6636 = Kaz 199}: the small northeastern component of this pair
(outside the SDSS footprint) has been catalogued as a
Seyfert galaxy. We are unaware of a previously published spectrum of the AGN.
In our new data, the strong lines would classify both galaxies as star-forming, but 
[O I] has intensity $\approx 0.05 \times$ H$\alpha$ across a large region, at the very
high end of values seen in disc H II regions and potentially a sign of 
composite ionization sources. In our [O III]
images, the small galaxy (with reported AGN) is surrounded by an arc of emission-line knots
(Fig. \ref{fig-kaz199o3}),
in addition to the region tracing the tidally distorted arms of the larger galaxy. The sizes of
these knots are consistent with unresolved H II regions, rather than the filamentary appearance
often seen in giant AGN-ionized clouds (\citealt{mnras2012}, \citealt{Keel2015}). 

\begin{figure} 
\includegraphics[width=68.mm,angle=270]{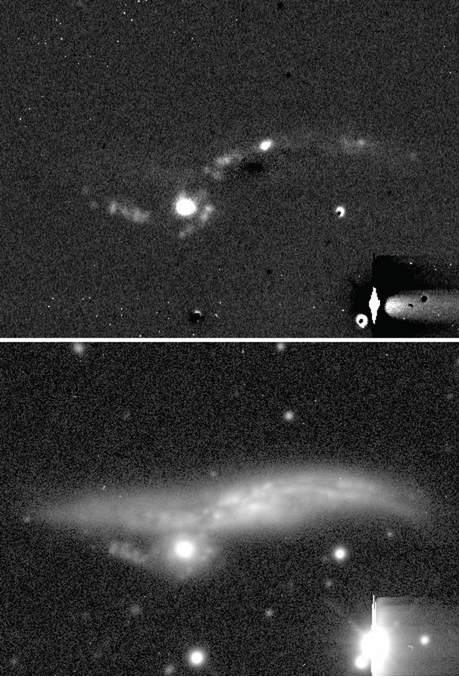} 
\caption{$V$-band continuum (left) and subtracted [O III] (right) images of Kaz 199 = NGC 6636
from the KPNO 2.1m telescope. A saturated star produced artifacts at lower left, and the 
colour changes associated with the dust lane in front of the bulge give negative residuals near the 
nucleus of the larger galaxy. The reported AGN is in the small companion galaxy to the
northeast (upper left). The field in each panel is $ 86 \times 119 $\arcsec\
showing north to the top and east to the left.}
 \label{fig-kaz199o3}
\end{figure}

Some systems attract attention because of what we {\it don't} see. The best example is
SDSS J084742.44+344504.4, also catalogued as PG 0844+349, a QSO with a nearby
large spiral companion galaxy projected 25\arcsec away (32 kpc). Its F$_{ion}$ and projected separation would be very favorable for 
cross-ionization, but our spectrum shows no evidence of this. 
Fig. \ref{fig-pg0844} shows the velocity structure of
this companion galaxy to the southwest, along its projected major axis. [O III] is detected
only at the two positions in its disc where the slit crosses the spiral arms, weaker than H$\alpha$, 
so there is no evidence of
external AGN ionization. An HST archival image (from Barth et al., program 9763) shows this 
to be a very symmetric, inclined grand-design spiral, similar in scale to the QSO host
galaxy itself (Fig. \ref{fig-pg0844hst}). The outer arms are so open in comparison to the inner structure that they
may be tidally induced; the dust lanes on the northern side of the arms plus our velocity measurements
show that the spiral pattern is trailing.
The slit spectroscopy by \cite{Hutchings} 
shows only weak emission in the southern spiral arm of the QSO host, without an obvious
enhancement as the slit crosses its western arc, so this feature more likely consists of
star-forming knots rather than AGN photoionization which fits as well with its very knotty structure compared to the
more diffuse and filamentary appearance of EELRs, as in \cite{Keel2017b}.
These data, and our results on the spiral companion, support either a picture of ionization cones directed
near the line of sight and out of the plane of the host galaxy, or one of intermittent high-luminosity
accretion phases, perhaps combined with a large physical separation between the pair members.
While there remains substantial uncertainty in the line-of-sight separation, this galaxy
pair may furnish an emission-line version of the same issues encountered in
comparing IGM ionization via the direct and transverse proximity effects (as summarized 
in the Introduction).

\begin{figure} 
\includegraphics[width=85.mm,angle=0]{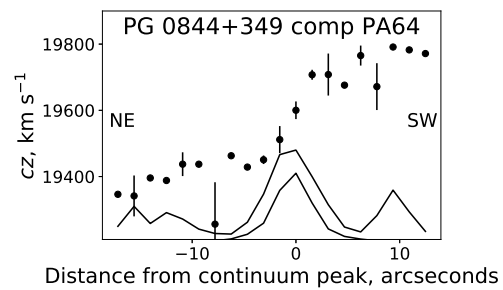} 
\caption{Spectroscopic results on the southern companion to PG 0844+349, shown as in
Fig. \ref{fig-longslit}. The traces across the bottom compare the continuum (lower, more compact) and (upper and more spatially extended) H$\alpha$ intensities along the slit.}
 \label{fig-pg0844}
\end{figure} 

\begin{figure} 
\includegraphics[width=68.mm,angle=-90]{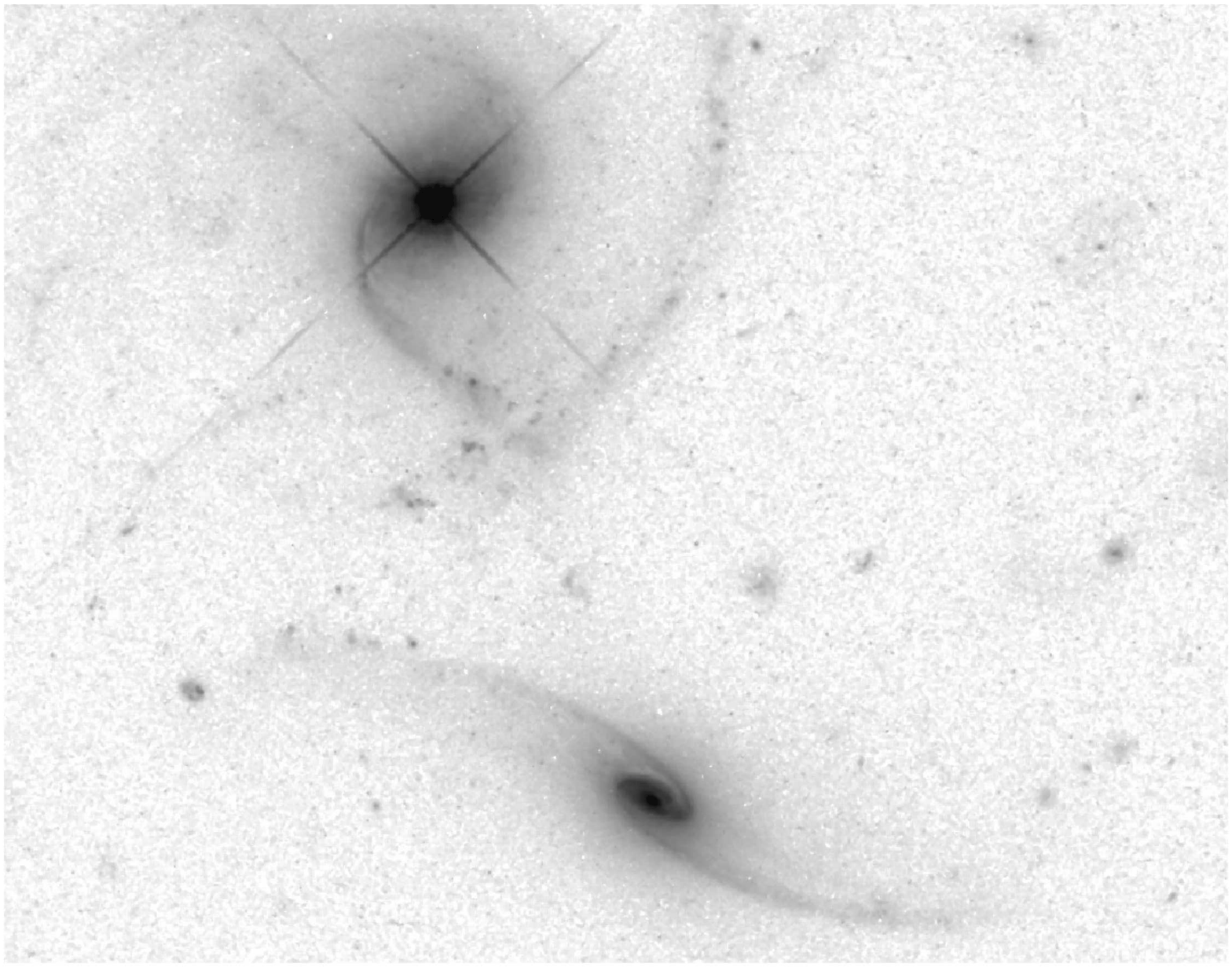} 
\caption{Archival  HST image of PG 0844+349 and companion galaxy to its south (from program HST-GO-9763, PI Barth). 
This is part of an ACS WFC image in the F625W filter, smoothed with a
3-pixel median for cosmetic reasons. The redshifts are a close match, with the companion 
at $z=0.0652$ compared to the luminous AGN at $z=0.0643$ for a projected velocity difference
near 270 km s$^{-1}$. As expected for trailing spiral
arms and the visible dust on the north side, the NE side of the disc is approaching. The area
shown spans $38 \times 49$ arcseconds with north at the top.}
 \label{fig-pg0844hst}
\end{figure} 

\subsection{Low-metallicity ``hidden" AGN-ionized gas, and }

Our primary tool for recognizing AGN-ionized gas is the set of strong emission-line ratios
in the ``classical" BPT diagram \citep{BPT}, based on [O III]/H$\beta$ and [N II]/H$\alpha$, incorporating
the revised class boundaries from \cite{Kewley2006}. These diagnostics become
less certain at lower metallicity as [N II] and [S II] become weaker. 
In this sample, for example, the BPT diagram would classify both SDSS J134203.48+183901.5 and SDSS J135255.67+252859.6
as star-forming galaxies, but broad H$\alpha$ in the first case, and strong and narrow He II $\lambda 4686$  in both, mark them as AGN.

\section{Conclusions}

Beginning with candidate pairs selected largely from the SDSS archive with the assistance of
Galaxy Zoo volunteer participants, we have surveyed a sample of close companion galaxies
to AGN 
for evidence of ionization of their interstellar medium by the external AGN (which we have termed
cross-ionization). Such cases
can help define the typical cone angles for escape of ionizing radiation, and the
typical radiative history of AGN over timescales $\approx 10^5$ years. Chosen on the
bases of predicted intensity of incident AGN radiation and fraction of solid angle around the AGN covered
by the companion, we observed the 32 most promising systems spectroscopically, of which 10 show
spectroscopic evidence for AGN ionization in the companion galaxy or tidal structures and are
candidates for cross-ionization.
This fraction (10/32=0.31) compares reasonably well with our simple geometric estimate $\approx 18$\% , based on
typical ionization cone opening angles $70^\circ$. A larger sample could, for example, place limits on
rapid precession of ionization cones, if it were faster than the local recombination timescales.

Since objects were selected for spectroscopy without prior knowledge of whether we would find evidence of
cross-ionization, we can compare the categories with and without such evidence with regard to our
selection parameters $r$ and $\theta$ to see whether any difference emerges, such as would be expected in
the simple case where the local intensity of isotropic AGN radiation controls the amount of excess ionization.
Among objects with spectra, the ranges and median values of F$_{\rm ion}$ are similar
for the subsets with and without evidence for cross-ionization; for example, the Kolmogorov-Smirnov test
suggests rejection of the hypothesis that the F$_{\rm ion}$ distributions are identical only at significance levels well below 90\%.
This could mean that preferential escape of ionizing radiation or long-term variability are indeed important in
determining the ionization environment of AGN. Our data cannot yet distinguish which might have the dominant role,
again parallel to studies of the Lyman $\alpha$ proximity effect (\citealt{OppenheimerSchaye},
\citealt{Khrykin}, \citealt{Schmidt2018}).

We illustrate this comparison, extended to both selection parameters, in Fig. \ref{fig-ioncompare}. The subsamples of our observed galaxies with and without 
signs of cross-ionization occupy the same regions of the ionizing flux-angle diagram, showing that these two parameters alone do 
not determine when this effect will be seen. The clustering of observed objects to the top and right reflects our observational priorities. 

\begin{figure} 
\includegraphics[width=85.mm,angle=180]{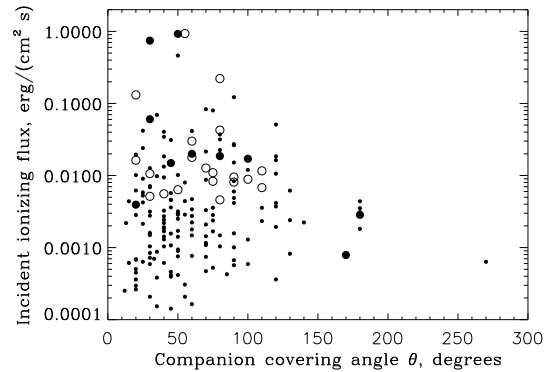} 
\caption{Comparison of subsamples with and without spectroscopic evidence for cross-ionization. The companion covering angle
$\theta$ is shown as in Fig. \ref{fig-schematic}, while the incident ionizing radiation from the AGN is estimated as in the text,
using the median projection correction from Monte Carlo simulations. Small filled circles show all objects in the initial finding list 
with ionizing flux above $10^{-4}$ erg cm$^{-2}$ s$^{-1}$ that we did not observe. Large filled circles indicate galaxies with 
evidence for cross-ionization
as in Table \ref{tbl-spectra}, while open circles show systems observed spectroscopically and not satisfying our criteria
for cross-ionization. }
 \label{fig-ioncompare}
\end{figure} 

The systems with the best-attested EELRs in our spectroscopic sample,  NGC 5278/9 and NGC 6081, each 
host dual AGN of roughly comparable luminosity, and highlight the ambiguities introduced in
such pairs. Both AGN may contribute to the extended ionization; untangling their
roles in the likely presence of structured obscuration will require more detailed information
on kinematics and the three-dimensional layout of each system.

At this point, our results largely provide a candidate list for detecting cross-ionization.
Further work in this direction could be pursued fruitfully with integral-field spectroscopy or
narrowband emission-line imaging. Such data could define the regions within which AGN
ionization is important. Emission-line imaging at high angular resolution would provide a more sensitive probe
for the role of external AGN in companion galaxies, 
allowing separation of the low-density diffuse ISM (whose ionization balance
would be more strongly altered by external radiation) from high-surface-brightness 
star-forming regions.

\section*{Acknowledgements}
This work would not have been possible without the contributions of citizen scientists as part of the Galaxy Zoo project. 
Among the Galaxy Zoo volunteers, particularly extensive contributions 
were made by users zutopian
and Blue\_Crew. Other 
contributors to the sample compilation were Alice, Blackprojects, Budgieye, 
ElisabethB, Hanny van Arkel, Jean Tate, Kiske, LankyYankee, Lightbulb50, 
LynnSeguin, Ranny44, egalaxy, elizabeth, fatha731, graham d, joinpep,
planetaryscience, and stellar190.

The work is partly based on observations obtained with the 6-m telescope
of the Special Astrophysical Observatory of the Russian Academy
of Sciences. The  analysis of ionized  gas in the systems NGC 5278/9 
and UGC 6081  was supported by the grant of Russian Science Foundation project 17-12-01335 ``Ionized gas in galaxy discs and beyond the optical radius".

Galaxy Zoo was made possible by funding from a Jim Gray Research Fund from Microsoft and The Leverhulme Trust. We thank the Lick Observatory staff for their assistance in obtaining
the data. VNB acknowledges assistance from National Science Foundation (NSF)
Research at Undergraduate Institutions (RUI) grant AST-1312296. Findings and conclusions
do not necessarily represent views of the NSF.

Funding for the creation and distribution of the SDSS Archive has been
provided by the Alfred P. Sloan Foundation, the Participating Institutions,
the National Aeronautics and Space Administration, the National Science
Foundation, the U.S. Department of Energy, the Japanese Monbukagakusho,
and the Max Planck Society. The SDSS Web site is http://www.sdss.org/. 
The SDSS is managed by the Astrophysical Research Consortium (ARC) for
the Participating Institutions. The Participating Institutions are The
University of Chicago, Fermilab, the Institute for Advanced Study, the Japan
Participation Group, The Johns Hopkins University, Los Alamos National
Laboratory, the Max-Planck-Institute for Astronomy (MPIA), the
Max-Planck-Institute for Astrophysics (MPA), New Mexico State
University, Princeton University, the United States Naval Observatory, and
the University of Washington.

This research has made use of the NASA/IPAC Extragalactic Database (NED),
which is operated by the Jet Propulsion Laboratory, Caltech, under
contract with the National Aeronautics and Space Administration. IRAF is distributed by the National Optical Astronomy Observatory, which is operated by the Association of Universities for Research in Astronomy (AURA) under a cooperative agreement with the National Science Foundation. The authors are honored to be permitted to 
conduct astronomical research on Iolkam Du'ag (Kitt Peak), a mountain with particular significance to the 
Tohono O'odham Nation.


\appendix

\section{Finding list: low redshift AGN/companion galaxy pairs}

For completeness, table \ref{tbl-allsources} gives relevant properties of all 212
AGN/galaxy pairs considered. Selection properties were projected angle subtended 
by the companion galaxy around the AGN ($\theta$) and predicted ionizing flux from the
AGN at the projected distance of the companion galaxy (F$_{\rm ion}$). F$_{\rm ion}$
was estimated as in Section 2.1, using the [O III] $\lambda 5007$ fluxes calculated,
where possible, from the SDSS spectra (based on tabulated line equivalent width EW in \AA\  and
local continuum flux F$_{\rm cont}$, as tabulated in SDSS DR7). Objects are identified both by coordinate designation, with initial J 
indicating designations used in the Sloan surveys \citep{SDSSDR12}; and, for SDSS objects, by the
Data Release 12  \citep{SDSSDR12} numerical ObjID for ease of data retrieval. For compactness,
F$_{\rm cont}$ is listed in units
$10^{-17}$ erg cm$^{-2}$ s$^{-1}$ \AA$^{-1}$. F$_{\rm ion}$ is given in 
arbitrary units, since the quantity has systematic uncertainties as well as projection effects and was used only
to rank objects for spectroscopic observation.

In a few cases we used additional sources for emission-line data: \cite{Petrov85} for NGC 6786, \cite{Heckman84} for the
companion of PG 1048+342, and \cite{KKHH85} for NGC 5278/9.

``Type" indicates the AGN type - 1 for broad-line objects (as in Sy 1), 2 for narrow-line objects, and 0 for BL Lac objects.

\begin{table*}
\caption{Finding list.}
\label{tbl-allsources}
\begin{tabular}{lccccrrrrrl}
\hline
Coord ID & SDSS ObjID & Type & $z$(AGN) & $z$(compn) & r\arcsec & $\theta$\degr  & F$_{\rm cont}$ & [O III] EW & F$_{\rm ion}$ & Notes \cr
\hline
J001707.95+011506.2 & 1237678617419710603 & 2 & 0.1057 & 0.1074 & 11 & 40 & 19.90 & 32.10 & 22 & \cr
J002228.37-005830.6 & 1237657189833900309 & 1.5 & 0.1060 & 0.1060 & 12 & 70 & 9.00 & 25.80 & 6.8 & \cr
J002944.89+001011.1 & 1237663784200634427 & 2 & 0.0598 & 0.0594 & 5 & 30 & 28.7 & 114.3 & 552 \cr
J005754.03+012013.8 & 1237678617424166953 & 2 & 0.0567 &      & 9 & 30 & 325 & 402 & 6800 & UM 293 \cr
J011309.59+021716.5 & 1237680100234887247 & 1.9 & 0.0462 &  & 72 & 28 &  &  & & UGC 768  \cr
J011448.67-002946.0 & 1237666338653012001 & 1 & 0.0338 & 0.0349 & 17 & 45 & 52.00 & 12.90 & 9.8 & UGC 793 \cr
J011659.06+001933.3 & 1237666339727015947 & 1 & 0.0782 & 0.0077 & 9 & 55 & 19.20 & 2.80 & 2.8 &  \cr
J013037.75+131252.0 & 1237649918432116906 & 2 & 0.0722 & 0.0726 & 16 & 100 & 14.00 & 23.30 & 5.4 &  \cr
J020436.76-115943.4 & 1237676674463826080 & 1.5 & 0.0726 &  & 24 & 45 &  &  &  &   \cr
J021209.23+072127.5 &  1237670016730333231 & 2 & 0.1424 &  & 6 & 45 & 4.20 & 5.00 & 2.5 &  \cr
J022226.11-085701.3 & 1237652900226596935 & 1 & 0.1666 &  & 3 & 70 & 15.20 & 106.60 & 759 & \cr
J022909.70+010043.0 & 1237678617434128461 & 2 & 0.1283 & 0.1288 & 8 & 30 & 8.60 & 11.60 & 6.6 &  \cr
J024222.87-011009.0 & 1237660024521556048 & 2 & 0.0372 & 0.0376 & 23 & 20 & 17.20 & 17.80 & 2.4 &  \cr
J025445.35+010308.4 & 1237678437018108063 & 2 & 0.1367 & 0.1366 & 9 & 20 & 8.70 & 7.40 & 3.3 &  \cr
J025740.83-163046.0 & 1237667244870008884 & 1 & 0.0679 &  & 5 & 80 &  &  &  &  \cr
J030858.44-001549.0 & 1237660240312467582 & 2 & 0.2064 & 0.2069 & 3 & 80 & 5.50 & 80.00 & 206 &  \cr
J033213.20+001546.7 & 1237666301093675376 & 1 & 0.0861 & 0.0875 & 8 & 100 & 26.40 & 6.80 & 11.8 &  \cr
J072612.51+412233.7 & 1237673706650796616 & 1 & 0.1324 &  & 4 & 70 & 11.50 & 6.30 & 19.1 &  \cr
J073402.62+433236.9 & 1237663916261703871 & 1 & 0.0831 & 0.0823 & 8 & 45 & 23.00 & 15.00 & 23 & \cr
J073906.17+442409.9 & 1237663915725357374 & 2 & 0.1345 & 0.1408 & 4 & 80 & 12.50 & 4.10 & 13.5 & \cr
J074408.20+435935.6 & 1237663530254663877 & 2 & 0.1331 & 0.1348 & 5 & 70 & 7.00 & 9.00 & 10.6 & \cr
J074429.85+284721.9 & 1237657119477989769 & 2 & 0.1015 & 0.1025 & 4 & 110 & 12.00 & 10.30 & 33 &  \cr
J074547.87+265537.9 & 1237657630578376999 & 1.8 & 0.1148 &  & 8 & 40 & 29.00 & 88.10 & 168 &  	 \cr
J075433.27+204635.2 & 1237661087490441736 & 1.8 & 0.1146 & 0.1146 & 8 & 50 & 8.00 & 40.00 & 21 & \cr
J075648.78+501016.6 & 1237663915728175471  & 1 & 0.2366 & 0.2364 & 6 & 45 & 9.00 & 13.60 & 14.3 &  \cr
J075729.04+351105.9 & 1237654626788704563 & 2 & 0.1117 & & 3 & 110 & 16.1 & 14.1 & 106 & \cr
J080059.77+343412.0 & 1237674290219450784 & 2 & 0.0823 & 0.0811 & 8 & 40 & 20.00 & 22.60 & 30 &  \cr
J080435.27+363711.5 & 1237654652024390067 & 2 & 0.0913 &  & 11 & 40 & 15.00 & 22.80 & 11.9 &  \cr
J080514.10+364942.9 & 1237654652024521006 & 2 & 0.0841 &  & 16 & 20 & 14.00 & 11.60 & 2.7 &  \cr
J080647.76+282512.8 & 1237658192145744217 & 2 & 0.1426 &  & 8 & 75 & 10.70 & 6.80 & 4.8 &  \cr
J081411.26+144735.4 & 1237667253453521141 & 2 & 0.1572 &  & 9 & 30 & 9.50 & 10.88 & 5.4 &  \cr
J082017.99+465125.3 & 1237651250946244661 & 2 & 0.0524 &  & 8 & 50 & 12.00 & 29.00 & 23 &  \cr
J082034.78+153111.2 & 1237667253454241925 & 1 & 0.1438 &  & 8 & 90 & 14.30 & 6.50 & 6.1 & \cr
J082913.59+264009.9 & 1237661087494766877 & 2 & 0.0574 & 0.0571 & 18 & 20 & 26.00 & 50.00 & 16.9 & \cr
J083049.39+102712.5 & 1237671262269997354 & 2 & 0.0727 & 0.0720 & 17 & 33 & 18.20 & 24.20 & 6.4 & \cr
J083116.14+351707.4 & 1237657629510271237 & 1.9 & 0.1601 &  & 3 & 90 & 10.00 & 53.20 & 249 &  \cr
J083202.71+093759.1 & 1237671261196255490 & 2 & 0.0750 &  & 4 & 130 & 21.00 & 10.20 & 56 &   \cr
J083224.28+355135.9 & 1237657775539945655 & 2 & 0.1368 &  & 6 & 60 & 15.00 & 14.30 & 25 &   \cr
J083728.08+391723.7 & 1237657401878708344 & 2 & 0.1147 &  & 20 & 30 & 17.50 & 41.80 & 7.7 &\cr
J083848.14+040734.0 & 1237658423540580569 & 2 & 0.0476 & 0.0484 & 9 & 20 & 72.00 & 39.70 & 149 &  \cr
J083902.96+470756.3 &  1237654381974716589  & 2 & 0.0524 & 0.0534 & 18 & 20 & 38 & 71 & 36 \cr
J084135.08+010156.2 & 1237650797286850905 & 2 & 0.1106 & & 4 & 80 & 19.2 & 400.1 & 2020 &  \cr
J084742.44+344504.4 & 1237658191076720756 & 1 & 0.0640 & 0.0653 & 22 & 30 & 850 & 3.1 &  97 & PG 0844+349 \cr
J084810.11+351534.3 & 1237664871356891234 & 2 & 0.0573 & 0.0570 & 4 & 80 & 19.5 & 33.2 & 170 & \cr
J085152.62+522833.0 & 1237651190821617882 & 1 & 0.0645 &  & 11 & 70 & 14.00 & 20.40 & 9.9 & \cr
J085820.50+642048.4 &  1237663917882802416 & 1 & 0.1158 &  & 5 & 30 & 11.50 & 59.70 & 116 &\cr
J090128.46+352030.1 & 1237664871894941906 & 2 & 0.1053 &  & 7 & 40 & 30.10 & 46.90 & 121 &  \cr
J090134.48+180943.0 & 1237667429554585733 & 2 & 0.0665 & 0.0665 & 9 & 40 & 32.00 & 14.90 & 25 &  \cr
J090317.39+100100.5 & 1237661069779009763 & 2 & 0.0619 &  & 12 & 130 & 31.12 & 24.10 & 22 & UGC 4748 \cr
J090436.92+553602.9 & 1237651252024836118 & 1 & 0.0372 & & 12 & 75 & 80.0 & 32.3 & 76 & \cr
J090604.60+170120.0 & 1237667485918429442 & 2 & 0.1007 &  & 8 & 45 & 15.00 & 52.80 & 52 &  \cr
J090743.49+013327.9 & 1237660670347706621 & 1 & 0.1643 &  & 5 & 75 & 18.70 & 6.90 & 22 & \cr
J090810.33+500923.1 & 1237654653642801161 & 1 & 0.0981 & 0.1073 & 13 & 50 & 44.00 & 23.80 & 26 &  \cr
J091029.08+203401.9 & 1237667210510336112 & 2 & 0.0289 & 0.0280 & 25 & 270 & 75.00 & 11.40 & 5.8 &  \cr
J091141.65+370204.1 & 1237660635988492309 & 1 & 0.0865 & 0.0862 & 14 & 30 & 30.00 & 39.20 & 25 &  \cr
J091319.00+532958.4 & 1237654383052062933 & 1.9 & 0.2549 &  & 3 & 20 & 13.00 & 9.30 & 57 &  \cr
J091449.05+085321.1 & 1237660670347706621 & 1 & 0.1399 &  & 3 & 75 & 10.10 & 6.90 & 33 &  \cr
J091540.88+155639.7 & 1237667735026532578 & 2 & 0.1434 &  & 9 & 40 & 11.00 & 35.00 & 20 &  \cr
J091930.42+133258.7 & 1237671125373485313 & 1 & 0.2291 & 0.2297 & 8 & 120 & 55.00 & 9.40 & 34 &  \cr
J093044.15+084929.7 & 1237661064413249768 & 2 & 0.1304 & 0.1300 & 4 & 40 & 11.00 & 108.40 & 314 &  \cr
J093811.95+045356.5 & 1237658298454180052 & 1.8 & 0.1626 &  & 4 & 180 & 12.00 & 10.20 & 32 &  \cr
094542.04-141934.8 &  & 2 & 0.0077 & 0.0081 & 87 & 20 & &  &  & NGC 2992 \cr
J094716.12+534944.8 & 1237655108368597025 & 2 & 0.0383 &  & 15 & 90 & 16.20 & 47.60 & 14.4 &  \cr
\hline
\end{tabular}

\end{table*}

\begin{table*}
\contcaption{Source list}
\label{tbl-allsources:continued}
\begin{tabular}{lccccrrrrrl}
\hline
Coord ID & SDSS ObjID & Type & $z$(AGN) & $z$(compn) & r\arcsec & $\theta$\degr  & F$_{\rm cont}$ & EW & F$_{\rm ion}$ & Notes \cr
\hline
J094741.58+633939.2 & 1237651540315930774 & 2 & 0.1390 &  & 3 & 50 & 16.00 & 12.40 & 93 & \cr
J095847.90+144526.2 & 1237671261742694538 & 2 & 0.0767 &  & 7 & 60 & 15.50 & 10.10 & 13.5 &  \cr
J095958.22+030232.7 & 1237654599952760951 & 2 & 0.0907 & 0.0906 & 17 & 40 & 20.00 & 44.70 & 13.0 &  \cr
J100531.67+183939.0 & 1237667782284017791 & 2 & 0.0779 & 0.0780 & 7 & 70 & 14.00 & 10.90 & 13.1 &  \cr
J100534.70+392852.8 & 1237660772882907153 & 1 & 0.1408 &  & 12 & 25 & 22.00 & 9.00 & 5.8 &  \cr
J100602.50+071131.8 & 1237658300604743843 & 2 & 0.1205 & 0.1218 & 5 & 80 & 25.00 & 80.50 & 339 &  \cr
J101043.36+061201.4 & 1237658423550607494 & 1 & 0.0978 & & 4 & 20 & 39.3 & 116.1 & 1200 & \cr
J101202.59+301303.0 & 1237665097393963129 & 1 & 0.0498 &  & 12 & 70 & 55.00 & 49.60 & 80 &  \cr
J101439.55-004951.2 & 1237654669200326683 & 2 & 0.0491 & 0.0483 & 19 & 20 & 41.00 & 13.20 & 6.3 & UGC 5528 \cr
J101653.82+002857.0 & 1237654670811201558 & 2 & 0.1163 &  & 2 & 50 & 20.90 & 191.70 & 4220 & \cr
J101932.86-032014.9 & 1237650803195248686 & 1 & 0.0500 & 0.0492 & 37 & 35 &  &  &  & Mkn 1253  \cr
J102027.52+135530.1 & 1237661070861140238 & 2 & 0.1449 &  & 6 & 15 & 16.00 & 21.40 & 40 &  \cr
J102250.45+495936.5 & 1237657858213871668 & 2 & 0.1155 &  & 16 & 55 & 11.70 & 9.70 & 1.9 &  \cr
J102536.43+371317.1 & 1237664669509550195 & 2 & 0.0608 &  & 5 & 180 & 34.00 & 7.00 & 40 &  \cr
J102850.92+163917.9 & 1237671262819713088 & 2 & 0.1720 &  & 5 & 75 & 5.40 & 16.80 & 15.3 & \cr
J102940.71+015555.1 & 1237651753463578759 & 2 & 0.0412 & 0.0418 & 40 & 50 & 37.50 & 38.00 & 3.8 & UGC 5694 \cr
J103000.93+085202.6 & 1237660583909720086 & 2 & 0.0522 &  & 20 & 35 & 63.00 & 12.00 & 8.0 &  \cr
J103011.81+035410.7 & 1237654604787745026 & 2 & 0.0510 &  & 11 & 50 & 22.00 & 11.20 & 8.6 &  \cr
J103216.14+505120.0 & 1237657589242396682 & 1 & 0.1737 &  & 3 & 120 & 22.40 & 16.10 & 169 &  \cr
J103600.37+013653.5 & 1237651752927363244 & 2 & 0.1068 &  & 7 & 25 & 42.00 & 105.90 & 382 &  \cr
J103607.13+191048.1 & 1237667781213421682 & 2 & 0.1117 &  & 10 & 70 & 14.00 & 7.30 & 4.3 &  \cr
J103655.60+380321.4 & 1237662226206228529 & 2 & 0.0507 & 0.0500 & 29 & 40 & 48.60 & 23.20 & 5.6 & \cr
J103734.22+140120.5 & 1237661070326104198 & 2 & 0.2060 &  & 5 & 20 & 23.00 & 46.00 & 178 & \cr
J103819.51+161445.1 & 1237671261746888940 & 2 & 0.1359 &  & 3 & 35 & 13.00 & 14.70 & 89 &  \cr
J103825.16-002331.1 & 1237654669739819119 & 2 & 0.0963 &  & 4 & 180 & 14.00 & 4.50 & 16.6 &  \cr
J104232.05+050241.9 & 1237654602104897626 & 2 & 0.0272 & 0.0273 & 9 & 180 & 26.0 & 19.4 & 26 & NGC 3341 \cr
J104519.45+123839.3 & 1237661068179406916 & 2 & 0.0394 & 0.0398 & 8 & 60 & 45.00 & 14.50 & 43 &  \cr
J105030.47+232931.4 & 1237667538012602552 & 1.8 & 0.0604 & & 5 & 50 & 18 & 19.2 & 58 & \cr
J105128.41+335850.8 & 1237665129609691296 & 1 & 0.1828 &  & 5 & 55 & 9.00 & 17.30 & 26 &  \cr
J105143.89+335926.7 & 1237665129609756690 & 1 & 0.1671 & 0.1670 & 4 & 40 & 81.00 & 17.30 & 369 & PG 1048+342 \cr
J105203.60+060349.4 & 1237658422481387583 & 1.8 & 0.1281 & 0.1285 & 11 & 15 & 18.00 & 9.00 & 5.6 &  \cr
J105222.48+180711.1 & 1237668288550076449 & 1 & 0.1322 &  & 2 & 120 & 26.00 & 17.00 & 466 &  \cr
J105418.31+181344.7 & 1237668288550273059 & 2 & 0.0811 &  & 10 & 30 & 16.00 & 20.90 & 14.1 &  \cr
J105434.78+302211.7 & 1237665329844322346 & 2 & 0.1055 &  & 13 & 20 & 16.00 & 15.50 & 6.2 &  \cr
J105503.48+425136.8 & 1237661850925793371 & 2 & 0.0592 & 0.0590 & 11 & 50 & 35.00 & 22.40 & 27 & \cr
J105505.07+414944.8 & 1237661871322955907 & 2 & 0.1077 &  & 2 & 120 & 8.20 & 17.30 & 149 &  \cr
J105521.02+291849.3 &  1237667212668895318 & 2 & 0.1408 &  & 3 & 30 & 6.50 & 40.00 & 122 &  \cr
J105738.07+082007.5 & 1237671930135445716 & 2 & 0.0371 & 0.0366 & 26 & 60 & 15.00 & 7.90 & 0.7 &  \cr
J110017.99+100256.8 & 1237671932283322446 & 2 & 0.0360 & 0.0361 & 20 & 170 & 40.00 & 17.10 & 7.2 & UGC 6081  \cr
J110157.90+101739.3 & 1237658493883711561 & 1.5 & 0.0341 & 0.0347 & 31 & 80 & 48.3 & 198.5 & 42 & Tol 1059+105 \cr
J110301.26+120543.0 & 1237661950246322306 & 2 & 0.1184 &  & 6 & 45 & 16.00 & 22.40 & 42 &  \cr
J110427.30+381232.2 & 1237662224597909569 & 0 & 0.0302 & 0.0313 & 12 & 20 &  &  &  & Mkn 421 \cr
J110525.11+215229.6 & 1237667783364051073 & 2 & 0.1475 &  & 4 & 80 & 11.80 & 93.00 & 289 &  \cr
J110544.44+195746.3 & 1237668294982500461 & 2 & 0.1043 &  & 4 & 90 & 16.00 & 13.00 & 55 &  \cr
J110550.53+205113.3 & 1237667734501326907 & 2 & 0.1015 &  & 8 & 25 & 22.00 & 26.70 & 39 &  \cr
J110612.16+061347.6 & 1237658422482895003 & 2 & 0.0979 & 0.0986 & 17 & 20 & 12.00 & 23.20 & 4.1 &  \cr
J110639.11+433620.9 & 1237661872934682808 & 2 & 0.1185 & 0.1186 & 5 & 110 & 23.40 & 5.40 & 21 &  \cr
J110816.45+293210.5 & 1237667255082745989 & 2 & 0.0466 & 0.0467 & 9 & 60 & 25.00 & 12.50 & 16 &  \cr
J111118.04+352308.3 & 1237664817671438525 & 2 & 0.0250 &  & 9 & 45 & 45.00 & 14.10 & 33 &  \cr
J111159.48+013643.4 & 1237651752931360818 & 1.9 & 0.1108 & 0.1099 & 7 & 30 & 13.00 & 25.80 & 29 &  \cr
J111545.61+065143.2 & 1237661971178586191 & 1 & 0.1492 &  & 9 & 25 & 12.00 & 85.30 & 53 &  \cr
J111623.16+532413.3 & 1237657589782413395 & 2 & 0.1123 & 0.1125 & 2 & 100 & 17.00 & 6.10 & 109 &  \cr
J111943.34+451612.6 & 1237661852538568731 & 2 & 0.2036 & 0.2030 & 6 & 60 & 10.00 & 323.60 & 379 &  \cr
J112336.17+521810.0 & 1237657857681260680 & 2 & 0.1579 &  & 11 & 30 & 12.00 & 32.50 & 13.6 &  \cr
J112353.48+655133.5 & 1237651537636032650 & 2 & 0.1326 &  & 7 & 50 & 8.80 & 10.90 & 8.2 &  \cr
J112525.71+533344.4 & 1237658800960700483 & 2 & 0.0804 & 0.0808 & 17 & 50 & 26.00 & 18.30 & 6.9 &  \cr
J112602.45+343448.1 & 1237665025437139019 & 1.5 & 0.1113 & 0.1115 & 11 & 140 & 21.00 & 27.90 & 20 & \cr
J112949.25+172020.2 & 1237668585971253453 & 1 & 0.1571 &  & 4 & 45 & 12.00 & 41.60 & 132 & \cr
J113003.10+655629.3 & 1237651271361495049 & 1 & 0.1326 &  & 4 & 90 & 19.60 & 26.80 & 138 &  \cr
J113215.11+244516.8 & 1237667550885314686 & 1 & 0.1711 &  & 4 & 50 & 7.40 & 6.60 & 12.9 &  \cr
J113240.25+525701.3 & 1237657630599872600 & 2 & 0.0266 & 0.0271 & 34 & 30 & 180 & 71.4 & 47 & Mkn 176 \cr
J113321.04+373944.3 & 1237664819821019315 & 2 & 0.1754 &  & 6 & 30 & 7.90 & 10.30 & 9.5 &  \cr
J113323.97+550415.8 & 1237658802571968533 & 1 & 0.0085 & 0.0078 & 8 & $>120$ & 1.30 & 11.90 & 1.0 & Mkn 177 \cr
\hline
\end{tabular}
\end{table*}

\begin{table*}
\contcaption{Finding list}
\label{tbl-allsources:continued}
\begin{tabular}{lccccrrrrrl}
\hline
Coord ID & SDSS ObjID & Type & $z$(AGN) & $z$(compn) & r\arcsec & $\theta$\degr  & F$_{\rm cont}$ & EW & F$_{\rm ion}$ & Notes \cr
\hline
J113342.78+353135.4 & 1237665026511601678 & 1 & 0.2018 & 0.2036 & 4 & 90 & 31.00 & 29.50 & 241 &  \cr
J113443.24+153848.0 & 1237661070869069974 & 2 & 0.0698 & 0.0693 & 12 & 50 & 30.00 & 30.80 & 27 &  \cr
J113800.25+482623.2 & 1237658611445530792 & 1 & 0.1000 & 0.1002 & 4 & 120 & 15.00 & 24.50 & 97 &  \cr
J113858.89+141253.2 & 1237664289924055180 & 2 & 0.0805 & & 8 & 75 & 22.6 & 67.3 &  100 & \cr
J113942.52+315433.7 & 1237665368509710424 & 1.5 & 0.0089 & 0.0090 & 84 & 30 & 50.00 & 1.00 &  & NGC 3786  \cr
J114154.76+465656.9 & 1237660635464466499 & 2 & 0.0728 &  & 4 & 90 & 1.60 & 12.40 & 5.2 & \cr
J114252.83+325124.2 & 1237665369583779929 & 2 & 0.0666 & 0.0661 & 15 & 40 & 27 & 101.8 & 51 & \cr
J114719.94+075243.1 & 1237661972255736072 & 2 & 0.0826 & 0.0821 & 15 & 30 & 27.00 & 28.90 & 14.6 &  \cr
J115112.99+064105.9 & 1237661970645581860 & 1 & 0.0768 & 0.0807 & 13 & 80 & 25.00 & 15.30 & 9.5 &  \cr
J115535.35+125253.9 & 1237661813346009111 & 1 & 0.1216 &  & 8 & 25 & 57.00 & 5.20 & 19.5 & \cr
J120041.39+314746.2 & 1237665226229940236 & 2 & 0.1159 & & 4 & 55 & 80 & 400 & 8530 & \cr
J120149.74-015327.5 & 1237650762390765695 & 1 & 0.0907 & & 5 & 50 & 18.0 & 147.2 & 8426 & \cr
J120235.97+372157.8 & 1237664818749767887 & 2 & 0.1975 &  & 6 & 75 & 12.00 & 18.40 & 26 &  \cr
J120336.87+200629.2 & 1237668298202415270 & 2 & 0.2121 &  & 3 & 90 & 11.40 & 209.00 & 1115 &  \cr
J120349.20+020556.9 & 1237651753473867953 & 1.5 & 0.0812 &  & 8 & 40 & 23.50 & 10.90 & 17 & \cr
J120655.62+501737.4 & 1237658205047488538 & 1.5 & 0.0621 & 0.0633 & 50 & 35 & 35.50 & 22.90 & 1.4 &  \cr
J120712.57+165809.6 & 1237668623548022885 & 2 & 0.0723 & 0.0737 & 13 & 85 & 24.00 & 6.51 & 3.9 &  \cr
J120836.75+223934.3 & 1237667782833602610 & 1 & 0.1997 &  & 3 & 90 & 18.30 & 8.90 & 76 &  \cr
J121015.67+262837.4 & 1237667441896390838 & 2 & 0.1041 &  & 8 & 20 & 38.00 & 37.00 & 93 &  \cr
J121151.12+443327.6 & 1237661850395017428 & 2 & 0.1003 &  & 15 & 120 & 16.00 & 11.00 & 3.3 &  \cr
J121155.63+372113.4 & 1237664818750619663 & 2 & 0.0838 & 0.0834 & 8 & 90 & 14.00 & 41.30 & 38 &  \cr
J121303.35+513855.2 & 1237657855537053802 & 2 & 00849 & & 7 & 100 & 22.0 & 42.6 & 81 & \cr
J121418.25+293146.7 & 1237667253478424610 & 2 & 00632 & 0.0640 & 7 & 100 & 31. & 58.5 & 156 & Was 49 \cr
J121522.77+414620.9 & 1237661966355988618 & 1 & 0.1963 &  & 5 & 45 & 17.00 & 99.10 & 284 &  \cr
J121749.78+354449.6 & 1237665025978663019 & 2 & 0.0881 & 0.0880 & 4 & 120 & 14.00 & 10.30 & 38 &  \cr
J121943.13+132659.9 & 1237661813885436022 & 2 & 0.0647 & 0.0645 & 11 & 110 & 25 & 71 & 62 & \cr
J121953.10+554506.8 & 1237658918529925220 & 1 & 0.1075 &  & 11 & 25 & 19.50 & 70.90 & 48 & \cr
J122026.09+044631.9 & 1237655125009498294 & 2 & 0.0811 & 0.0803 & 45 & 12 & 8.00 & 200.00 & 3.3 &  \cr
J122157.16+080515.1 & 1237661972259537090 & 2 & 0.0718 &  & 12 & 45 & 30.20 & 47.60 & 42 &  \cr
J122500.80+014401.8 & 1237651752939290772 & 2 & 0.0897 & 0.0898 & 25 & 45 & 28.00 & 20 & 3.8 &  \cr
J122744.56-015140.2 & 1237650371558703219 & 1 & 0.1484 &  & 5 & 40 & 22.60 & 22.60 & 86 &  \cr
J123152.05+450443.1 & 1237661872941367322 & 1 & 0.0622 & 0.0621 & 20 & 60 & 55.00 & 12.00 & 7.0 &  \cr
J123351.61+195311.6 & 1237667915421778099 & 2 & 0.0737 & 0.0735 & 7 & 50 & 15.00 & 12.10 & 16 &  \cr
J123507.49+171509 & 1237668589724631143 & 2 & 0.0894 & 0.0898 & 11 & 40 & 19.30 & 26.00 & 17 & \cr
J124214.47+141147.0 & 1237662524694462694 & 2 & 0.1574 &  & 15 & 20 & 14.00 & 17.60 & 4.6 &  \cr
J124322.55+373858.0 & 1237664819290177687 & 2 & 0.0859 & 0.0865 & 6 & 90 & 37 & 17.3 &  74 & \cr
J124406.59+652925.5 & 1237654610679955615 & 2 & 0.1071 &  & 4 & 75 & 21.50 & 129.00 & 730 &  \cr
J130116.09-032829.0 & 1237650369414889508 & 2 & 0.0864 & 0.0867 & 11 & 70 & 34.80 & 24.90 & 30 &  \cr
J130240.41+514644.4 & 1237661958293028932 & 2 & 0.0539 & 0.0550 & 110 & 12 & 28.20 & 57.70 & 0.6 & UGC 8151 \cr
J130354.71-030631.8 & 1237650760786903297 & 2 & 0.0778 & & 6 & 60 & 36.2 & 38.6 & 163 & \cr
J130600.68+000125.0 & 1237648704582844527 & 2 & 0.1379 & 0.1377 & 13 & 20 & 11.00 & 94.60 & 26 &  \cr
J131135.65+142447.1 & 1237662525234479221 & 1 & 0.1140 &  & 9 & 45 & 60.00 & 42.10 & 131 &  \cr
J131318.92+504001.7 & 1237661957220139201 & 2 & 0.1585 &  & 6 & 55 & 13.20 & 26.10 & 40 &  \cr
J131517.26+442425.5 & 1237661851473739819 & 2 & 0.0355 & 0.0355 & 37 & 90 & 110.00 & 27.20 & 9.2 & \cr
J131639.74+445235.0 & 1237661852010676264 & 2 & 0.0906 & 0.0915 & 12 & 25 & 45.00 & 166.50 & 219 &  \cr
J131854.19+221825.5 & 1237667783377551471 & 2 & 0.1818 &  & 3 & 90 & 8.20 & 11.60 & 45 & \cr
J132751.63+330638.1 & 1237665127476428926 & 2 & 0.1412 &  & 2 & 70 & 13.70 & 13.60 & 196 &  \cr
J133920.85+393224.3 & 1237662305112293614 & 2 & 0.2240 &  & 3 & 90 & 39.00 & 200.00 & 3650 &  \cr
J134143.75+554025.5 & 1237661387065786478 & 2 & 0.0251 & 0.0253 & 38 & 45 & 3100.00 & 15.00 & 136 & NGC 5278/9  \cr
J134203.48+183901.5 & 1237668271899738298 & 1.8 & 0.0845 & & 5 & 80 & 17.6 & 134.4 &  399 & \cr
J134234.31+191338.9 & 1237668272436674636 & 1 & 0.0866 & 0.0866 & 9 & 180 & 49.00 & 11.10 & 28 &  \cr
J134730.70+603743.1 &  1237655369829580934 & 1 & 0.1433 & 0.1438 & 5 & 30 & 15.20 & 25.40 & 65 &  \cr
J134736.39+173404.6 & 1237668316456288331 & 2 & 0.0447 & 00451 & 10 & 90 & 40 & 51.9 & 87 & \cr
J134955.48+084039.6 & 1237662236939256056 & 2 & 0.0670 &  & 10 & 30 & 220.00 & 6.80 & 63 & \cr
J135255.67+252859.6 & 1237665531709554737 & 1 & 0.0636 &  & 5 & 60 & 40.00 & 40.70 & 274 &  \cr
J135429.05+132757.2 & 1237662529526890706 & 2 & 0.0633 & & 9 & 60 & 8.4 & 420 & 183 & \cr
J135713.88+115951.0 & 1237662199886184522 & 2 & 0.0205 & 0.0205 & 85 & 30 & 97.30 & 4.60 & 0.3 &  \cr
J140944.19+100429.7 & 1237662239088836793 & 2 & 0.0958 & 0.0951 & 26 & 12 & 12.40 & 30.40 & 2.3 &   \cr
J141041.49+223337.0 &  1237667911678099640 & 1 & 0.1725 &  & 3 & 35 & 31.50 & 43.00 & 634 &  \cr
J141104.81-001906.6 & 1237648704053051635 & 2 & 0.1387 & 0.1390 & 5 & 60 & 11.50 & 16.00 & 31 &  \cr
J141447.15-000013.1 & 1237648721245503521 & 2 & 0.0475 & 0.0474 & 5 & 70 & 24.0 & 28.8 & 116 & \cr
J141946.07+650353.2 & 1237674479196242084 & 1 & 0.1479 &  & 4 & 70 & 25.00 & 50.00 & 329 &  \cr
\hline
\end{tabular}
\end{table*}

\begin{table*}
\contcaption{Finding list}
\label{tbl-allsources:continued}
\begin{tabular}{lccccrrrrrl}
\hline
Coord ID & SDSS ObjID & Type & $z$(AGN) & $z$(compn) & r\arcsec & $\theta$\degr  & F$_{\rm cont}$ & EW & F$_{\rm ion}$ & Notes \cr
\hline
J142255.49-001711.4 & 1237648704054362331 & 2 & 0.1302 &  & 5 & 40 & 5.00 & 200.00 & 168 &  \cr
J142755.36-003341.8 & 1237648720710074546 & 2 & 0.0795 & 0.0792 & 7 & 50 & 15.00 & 22 & 29 &  \cr
J142802.22+062619.5 & 1237655744015368412 & 2 & 0.1155 &  & 13 & 45 & 8.90 & 18.80 & 4.2 &  \cr
J143454.21+334934.5 & 1237662661597659257 & 2 & 0.0578 & 0.0587 & 14 & 35 & 30.00 & 5.4 & 3.5 &  \cr
J143515.65+023221.6  & 1237651754564190252 & 1 & 0.3048 &  & 6 & 25 & 46.70 & 15.10 & 82 &  \cr
J144213.96+231844.6 & 1237665532251406538 & 1 & 0.1830 &  & 9 & 30 & 6.10 & 5.90 & 1.9 &  \cr
J144248.27+120040.3 & 1237662529532133399 & 0 & 0.1631 &  & 2 & 120 & 42.00 & 0.40 & 18 &  \cr
J144733.17-013844.3 & 1237655499209310386 & 2 & 0.0428 & 0.0427 & 73 & 26 & 28.70 & 3.90 & 0.1 & \cr
J144937.71+184159.5 & 1237667783386988590 & 2 & 0.0321 & 0.0320 & 29 & 45 & 89.40 & 3.00 & 1.3 & \cr
J145002.97+255441.2 & 1237665178450788549 & 1.8 & 0.0604 &  & 7 & 60 & 15.00 & 15.80 & 20.4 &  \cr
J145051.51+050652.1 & 1237651822714028249 & 2 & 0.0275 & 0.0282 & 21 & 55 & 100.00 & 76.40 & 73 & NGC 5765  \cr
174659.8 +683637 &  & 1 & 0.0630 & 0.0623 & 7 & 120 &  &  &  & Kaz 163 \cr
182205.11+663719.2 &  & 2 & 0.0155 & 0.0146 & 24 & 150 &  &  &  & Kaz 199 NGC 6636 \cr
191152.98+731929.5 &  & 2 & 0.0250 & 0.0252 & 85 & 10 &  & 13.30 &  & NGC 6786   \cr
J220041.34+103308.0 & 1237678859532566564 & 2 & 0.0266 &  & 3 & 120 &  &  &  & Mkn 520  \cr
J220555.01-000755.2 & 1237663543146643720 & 2 & 0.0945 & 0.0939 & 13 & 60 & 13.30 & 4.40 & 1.5 &  \cr
J220701.98+101400.8 & 1237678876177072140 & 2 & 0.0266 & 0.0266 & 13 & 75 &  &  &  & NGC 7212 \cr
J225057.27-085410.8 & 1237652600111366327 & 1.8 & 0.0650 & 0.0644 & 6 & 30 & 35.00 & 23.50 & 96.3 &  \cr
J225452.22+004631.3 & 1237663445442691192 & 1 & 0.0909 & 0.0912 & 11 & 40 & 57.50 & 7.70 & 15.4 &  \cr
J225823.29+001603.8 & 1237663444906213507 & 2 & 0.1542 & 0.1543 & 10 & 45 & 16.10 & 5.30 & 3.6 & \cr
J230315.61+085226.0 &  1237679034548551752 & 1 & 0.0163 & 0.0160 & 71 & 50 & 0.00 &  & 0.0 & NGC 7469  \cr
J230443.47-084108.6 & 1237653438155128872 & 1 & 0.0470 & 0.0473 & 63 & 13 & 172.00 & 109.80 & 20 & Mkn 926  \cr
J232756.70+084644.3 & 1237679034014367754 & 2 & 0.0289 & 0.0295 & 45 & 80 &  &  &  & NGC 7674  \cr
J234428.33+010829.7 & 1237656909037699193 & 2 & 0.1206 & 0.1167 & 11 & 40 & 13.90 & 19.60 & 9.5 &  \cr
\hline
\end{tabular}
\end{table*}

\label{lastpage}

\end{document}